# µSR and inelastic neutron scattering investigations of the caged type Kondo semimetals: CeT$_2$Al$_{10}$ (T=Fe, Ru and Os)

D.T. Adroja[1,4$], A.D. Hillier[1], Y. Muro[2], T. Takabatake[3], A.M. Strydom[4], A. Bhattacharyya[1,4], A. Daoud-Aladin[1] and J. W. Taylor[1]

[1]ISIS Facility, Rutherford Appleton Laboratory, Chilton, Didcot Oxon, OX11 0QX, UK

[2]Liberal Arts and Sciences, Faculty of Engineering, Toyama Prefectural University, Imizu 939-0398, Japan

[3]Department of Quantum matter, ADSM, and IAMR, Hiroshima University, Higashi-Hiroshima, 739-8530, Japan

[4]Physics Department, University of Johannesburg, PO Box 524, Auckland Park 2006, South Africa

(Dated: 25th Nov. 2013)

Recently the Ce-based caged-type compounds having general formula CeT$_2$Al$_{10}$ (T=Fe, Ru and Os) have generated considerable interest due to the Kondo semiconducting paramagnetic ground state (down to 40 mK) observed in CeFe$_2$Al$_{10}$ and anomalously high magnetic ordering temperature with a spin gap formation at low temperatures in Kondo semimetals CeRu$_2$Al$_{10}$ and CeOs$_2$Al$_{10}$. The formation of long-range magnetic ordering out of the Kondo semiconducting/semimetallic state itself is extraordinary and these are the first examples of this enigmatic coexistence of electronic ground states. These compounds also exhibit strong anisotropy in the magnetic and transport properties, which has been explained on the basis of single ion crystal electric field (CEF) anisotropy in the presence of strongly anisotropic hybridization between localised 4f-electron and conduction electrons. Further they also exhibit a remarkable modification of magnetic and transport properties with doping on Ce, or T or Al sites. In this review we briefly discuss the bulk properties of these compounds, giving a detailed discussion on our muon-spin-relaxation (µSR) investigations and inelastic neutron scattering (INS) results. We present µSR and INS results of Ce(Ru$_{1-x}$Fe$_x$)$_2$Al$_{10}$ and CeOs$_2$Al$_{10}$ as well as the µSR results of NdFe$_2$Al$_{10}$, NdOs$_2$Al$_{10}$ and YFe$_2$Al$_{10}$ for comparison. The zero-field µSR spectra clearly reveal coherent two-frequency oscillations at low temperatures in CeT$_2$Al$_{10}$ (T=Ru and Os) and Ce(Ru$_{1-x}$Fe$_x$)$_2$Al$_{10}$ (x=0.3 to 0.5), which confirms the long-range magnetic ordering with a reduced moment of the Ce. On the other hand the µSR spectra of Ce(Ru$_{1-x}$Fe$_x$)$_2$Al$_{10}$ (x=0.8 and x=1) down to 1.2 K and 0.04 K, respectively exhibit a temperature independent Kubo-Toyabe (KT) term confirming a paramagnetic ground state. INS measurements on CeT$_2$Al$_{10}$ (T=Ru and Os) exhibit sharp inelastic excitations at 8 and 11 meV at 5 K due to an opening of a gap in the spin excitation spectrum. A spin gap of 8-12 meV at 7 K, with a strong Q-dependent intensity, is observed in the magnetic ordered state of Ce(Ru$_{1-x}$Fe$_x$)$_2$Al$_{10}$ with x=0.3 and 0.5 which remarkably extends into the paramagnetic state of x=0.8 and 1. The observation of a spin gap in the paramagnetic samples (x=0.8 and 1) is an interesting finding in this study and it challenges our understanding of the origin of the semiconducting energy gap in $CeT_2Al_{10}$ (T=Ru and Os) in terms of a hybridization gap opening only a small part of the Fermi surface, gapped spin waves, or a spin-dimer gap. Further, the µSR study of $NdFe_2Al_{10}$ below $T_N$ exhibits a clear sign of two frequency oscillations, which are absent in $NdOs_2Al_{10}$. Moreover the µSR study of $YFe_2Al_{10}$, which has been proposed as a compound exhibiting ferromagnetic critical fluctuations did not reveal any clear sign of critical magnetic fluctuations down to 60 mK, within ISIS µSR time window, which is unexpected for a T→0 quantum phase transition (QPT).

## Contents

IV. Conclusions and future work

# I. Introduction

In recent years the investigation of intermetallic compounds of rare-earth elements (with localised $4f^n$ electrons) and transition metals (with itinerant $d^n$ electrons) has been a frontier topic of research in magnetism due to the fact that several compounds of this family can be used as permanent magnets of outstanding quality, giant magnetocaloric materials, memory storage, optical sensor devices, and solid state thermoelectric coolers [1-7]. Not only the magnetic attributes of rare-earth metals and alloys, but also a further diversity of thermodynamic and structural features have attracted attention to some of their exceptional properties with respect to reversible absorption of hydrogen gas at room temperature and nearly atmospheric pressures providing ideal candidates for energy storage materials [7].

Besides the technological applications of intermetallic compounds, Ce and Yb-based compounds exhibit a rich fundamental physics due to localised $4f^n$ (n=1 for Ce and 13 for Yb) electrons, which are often predisposed towards a localised versus delocalised or itinerant behaviour [8-12]. Due to this unique property the compounds exhibit interesting behaviour such as the Kondo effect [13], in which local moment is screened by the conduction electrons cloud, heavy electron behaviour (or heavy fermion, HF) [14], in which the effective mass ($m^*$) of an electron becomes 1000 times or more heavier than the mass of a bare electron ($m_0$), mixed valence behaviour [15], in which the $4f^n$-electrons gain energy from the hybridization between conduction electrons to jump into the conduction band and vice-versa, reduced magnetic moment ordering [16], Kondo insulator or Kondo semiconductor [17-19], spin and charge gap formation [20], charge and spin density waves, metal-insulator transition, unconventional superconductivity [21-25], spin-dimer formation, non-Fermi-liquid (NFL) behavior and quantum criticality associated with quantum phase transitions (QPT) [26-33]. These phenomena arise due to the presence of strong hybridization between localised 4f-electrons and conduction electrons and are present mainly in Ce and Yb systems, due to the proximity of 4f-level close to the Fermi level, and occasionally also in Pr, Sm, Eu and Tm based compounds. It is interesting to note that these compounds can have two static valence states ($4f^n$ and $4f^{n+1}$) or dynamic valence fluctuations between the two valence states with a characteristic time scale, $t_{vf}$. Thus the experimental techniques, which have probing time faster than $t_{vf}$ will see two valence states separately (i.e. X-ray photoelectron spectroscopy (XPS) and X-ray absorption near edge structure (XANES)) [34]. While neutron scattering as well as most other bulk physical properties having probing time slower than $t_{vf}$ will give information about an average valence state [35].

In heavy fermion systems the existence of strong electron-electron correlations gives a huge electronic density of states near the Fermi level ($E_F$) (Fig.1a), which is responsible for the origin of heavy effective mass. Among the novel properties mentioned above, the fascinating the Kondo insulator behaviour observed in d- and f-electron systems has been of recurring interest to this community and has recently attracted considerable interest in theoretical and experimental condensed matter physics [17-19, 36-42]. Some of these materials exhibit a very small energy gap (Fig.1b), called a hybridization gap, near $E_F$ and it is believed that the gap arises in the lattice, from the hybridization between the localized electrons (d- or f-electrons) and the conduction electrons [28-29, 34-36]. The main theoretical interest in these materials is due to the existence of large many-body renormalizations [43].

The application of pressure, magnetic field or doping that acts an adjustable tuning parameter ($\delta$) in HF systems, as there is a strong competition between electronic localisation and itinerant behaviour, which provides an opportunity to continuously modify the properties of HF systems (Fig.2). Generally, the localization of the electrons leads to a fundamental state of antiferromagnetic (or in some cases ferromagnetic) nature, while their delocalization is accompanied by a paramagnetic Fermi liquid (FL) state, which is a key ingredient of our fundamental understanding of metals, and founded upon the Landau FL theory [44]. At low temperatures FL systems show characteristic behaviour with resistivity, $\rho(T) \propto T^n$ with n=2, linear heat capacity with temperature $C_v(T) \propto \gamma_{ele}T$ and temperature independent susceptibility, $\chi(T)$ [44]. The FL model is the correct description of the low-temperature measurable parameters of a metal provided that the electron interactions as T→0 become temperature independent and are short ranged in both space and time. Deviations from the FL behaviour has been observed at low temperatures in many HF systems, near a critical value of the tuning parameter ($\delta_c$), which is referred to as non-Fermi-liquid behaviour (NFL) [26, 27, 30, 31, 45]. NFL systems at low temperatures exhibit $\rho(T) \propto T^n$ with n~1, $C_v(T) \propto -\log(T)$ and $\chi(T) \propto -\log(T)/T$, which are expected for two-dimensional antiferromagnetic spin fluctuations [26, 45,46]. The phase diagram of HF systems, shown in Fig. 2, originates from the competition between the onsite Kondo interaction ($T_{Kondo}$), which destabilizes magnetism and support formation of nonmagnetic singlet, and intersite Ruderman-Kittel-Kasuya-Yosida (RKKY) interactions ($T_{RKKY}$), which favors a magnetic ground state [47]. The Kondo temperature varies exponentially with the product of the electron density of state $D(E_F)$ at $E_F$ and the exchange coupling $J_{sf}$ between conduction electrons and local moments, $T_{Kondo} \sim (D(E_F)* J_{sf})^{1/2}*\exp(-1/|D(E_F)* J_{sf}|)$, while magnetic RKKY interaction energy varies as square of this product, $T_{RKKY} \sim (D(E_F)* J_{sf})^2$. Doniach [47] made the first theoretical study of a Kondo lattice using a one dimensional chain of spins and this result is still used to classify many HF compounds on the phase diagram. It can be controlled via an adjustable parameter δ (called lattice density), which allows us to tune systems from an antiferromagnetic phase to a paramagnetic and Fermi-liquid regime. These two regimes are separated by a $T=0$ quantum phase transition (QPT), also called quantum critical point (QCP) [26, 46, 48]. The phase space near quantum phase transitions has been a source of new and unexpected behaviour in physical properties of solids. Among these are collective NFL behaviour and the development of an unconventional superconductivity. The notion of critical magnetic fluctuations is at the heart of HF superconductivity. The existence of magnetically mediated superconductivity in HF compounds could help to shed light on the question of whether magnetic interactions are relevant for describing the superconducting and normal-state properties of other strongly correlated electron systems, perhaps including the high-temperature copper oxide superconductors (HTSC) [49-50]. Although superconducting transitions temperatures (Tc) are remarkably different in HFSC (below 2-3K) and in HTSC (50-150K), these two families of materials have surprisingly similar phase diagrams and thus also the likelihood of a common origin of the superconductivity. The common behaviour can be understood by considering the ratio between $T_C$ and the electronic band width or Fermi temperatures.

## A. General aspects of caged- type structures

Compounds having a caged type crystal structure have been of enduring interest for example clathrates [51], filled skuttrudites [5, 52, 53], β-pyrochore [54], $RT_2X_{20}$ (R=rare earths T=transition metals and X=Al and Zn) [55], $R_3T_4Sn_3$ and many more [56-58]. Their crystal structure commonly possesses three-dimensional skeletons containing large atomic cages, inside of which relatively small atoms may naturally be located or otherwise may often be injected and these can rattle with large atomic excursions owing to either a virtual size mismatch or due to very weak chemical bonding as a consequence of large interatomic spacing. The rattling atoms thus have weak structural coupling, and strong electron–phonon (rattler) coupling, leading to a significant anharmonicity for rattling vibration for the central atom. These compounds have attracted a huge interest because rattling vibration may suppress the phonon part of the thermal conductivity, resulting in conditions favourable for the enhancement of thermoelectric efficiency [5, 52, 58]. Besides their suitability for exploiting the thermoelectric effect, the caged compounds of Ce, Pr and Yb are deemed to be especially suitable for thermoelectric power generation due to the high Seebeck coefficient found in strongly correlated compounds of these elements (the Seebeck coefficient features in the thermoelectric figure of merit as a squared quantity). These compounds are important moreover to investigate superconductivity, Kondo effect and anisotropic hybridization, quadrupolar ordering and many more exotic properties [55, 57, 58].

Recently the CeT$_2$Al$_{10}$ (T=Fe, Ru and Os) compounds, having caged-type orthorhombic crystal structure (Fig.3), have attracted considerable interest in strongly correlated electron systems' community, both experimentally and theoretically, due to the remarkable physical properties they exhibit [59-70]. Among the explanations forwarded to date to capture the perplexing physics found in these compounds, are for example the opening of a spin and charge gap, anisotropic hybridization and charge density modulation [59-60, 65-68]. The Ru and Os compounds order antiferromagnetically at $T_N$ = 27 and 28.5 K, respectively, while the Fe compound remains paramagnetic down to 50 mK [59-61]. It is interesting to note that the ordering temperature of these compounds is extraordinarily high compared to their isostructural Gd-compound, GdRu$_2$Al$_{10}$ ($T_N$ = 16 K) and GdOs$_2$Al$_{10}$ ($T_N$ = 18 K) (See Fig.4). This inconsistency becomes even more significant when noting that $T_N$ of CeRu$_2$Al$_{10}$ and CeOs$_2$Al$_{10}$ compounds is higher than most magnetically ordered Ce-compounds, with only a few exceptions (see Table. I). For example CeRh$_3$B$_2$ orders ferromagnetically (FM) with $T_C$=120 K [71, 72], while isostructural GdRh$_3$B$_2$ orders FM with $T_C$=90 [73]. This unusual behaviour is not expected based on the de Gennes scaling [74], which shows that the ordering temperature of an isostructural family of rare-earth compounds with different rare-earth ions across the series from Ce to Yb should scale with ~ $(g_J - 1)^2$ $J(J+1)$, where $g_J$ is the Lande g-factor and J is the total angular momentum. Thereby, the family member containing Gd should have the highest ordering temperature since it has the highest values of the de Gennes factor. Hence, the ordering temperature of the Ce-compounds should be a factor 0.0113 smaller than that of the Gd-compounds, which is not the case for most of the ordered Ce-compounds (see Table-I). This discrepancy can be partly understood by considering two facts. One is the exchange interactions, *J(Q)*, where Q is the wave vector for which *J(Q)* is maximum, not having a constant values across the series ($T_C$ (or $T_N$) ~ $J(Q)*(g_J - 1)^2$ $J(J+1)$). The other is the crystal field effect, which may change the ground state magnetic moment [75-76]. Usually rare-earth compounds are classified into exchange dominated or crystal-field-dominated systems. This classification is clearly not valid for CeRh$_3$B$_2$ where all energy scales (i.e. spin-orbit, crystal field, and exchange) must be taken into account to understand the ground state and low-lying magnetic excitations [77]. The spin wave excitation spectrum evidences high exchange interaction along the *c*-axis about two orders of magnitude higher than the ones in the basal plane of CeRh$_3$B$_2$. The easy axis is in the basal plane of the hexagonal structure and the saturation magnetization 0.4$\mu_B$ is strongly reduced compared to free cerium ion value 2.14 $\mu_B$. The electronic structure calculations of CeRh$_3$B$_2$ reveal that the unconventional ground state is stabilized by the strong 4f–4f direct mixing between the neighbored Ce atoms along the extremely small distance along the c-axis in the hexagonal crystal cell [77].

In this review we first discuss the structural properties and then magnetic and transport properties of CeT$_2$Al$_{10}$ and doped compounds. We then introduce μSR results on Ce(Ru$_{1-x}$Fe$_x$)$_2$Al$_{10}$, CeOs$_2$Al$_{10}$ followed by data on the local 4f-moment systems NdT$_2$Al$_{10}$ (T=Fe and Os), and we conclude with μSR findings on quantum critical system YFe$_2$Al$_{10}$. We present the inelastic neutron scattering investigations on Ce(Ru$_{1-x}$Fe$_x$)$_2$Al$_{10}$ and CeOs$_2$Al$_{10,}$ compared with published results. Finally, we summarize the present work with important conclusions and propose future investigations to be carried out for further understanding the complex behaviour of CeT$_2$Al$_{10}$ compounds.

## II. Experimental details

The polycrystalline samples investigated in the present work were prepared by argon arc melting and the details can be found in refs [59-61]. For the zero-field (ZF) µSR experiments, the powdered samples (thickness ~1.5mm) were mounted onto a 99.995+% pure silver plate using GE-varnish and were covered with 18 micron silver foil. We used the MuSR spectrometer in longitudinal geometry at the ISIS Pulsed Neutron and Muon Source, UK. At the ISIS facility, a pulse of spin polarised muons is produced every 20 ms and has a full-width at half-maximum (FWHM) of ~70 ns. The polarized muons are implanted into the sample and decay with a half-life of 2.2 µs into a positron which is emitted preferentially in the direction of the muon spin axis. Due to the magnetic moment of muons, the polarized muons precesses around a local (non-zero) magnetic field B$_{loc}$, with a precession frequency ω$_µ$ proportional to B$_{loc}$, ω$_µ$ = 2πf$_µ$=γ$_µ$B$_{loc}$, where f$_µ$ is the muon precession frequency and γ$_µ$ (=851.6 Mrad/s/T or 135.6 MHz/T) is the gyromagnetic ratio (γ$_µ$) of muon: 1MHz = 73.8 G. In a constant local field, therefore, moment of muon rotates by =γ$_µ$B$_{loc}$* t in the elapsed time t. As the typical data collection time window at the ISIS muon facility is 15µs, this permits detection of a local field as small as 0.04mT (or 0.4G). The emitted positrons are detected and time stamped in the detectors which are positioned before, F, and after, B, the sample. From the measured positron counts in the F and B detectors, N$_F$(t) and N$_B$(t), respectively, the asymmetry of the muon decay, G$_z$(t) is determined using

$$G_z(t)=(N_F(t)-\alpha N_B(t))/(N_F(t)+ \alpha N_B(t)) \quad (1)$$

where α is a calibration coefficient.

The inelastic neutron scattering measurements on the polycrystalline samples were carried out using the MARI/MERLIN time-of-flight (TOF) chopper spectrometers at the ISIS Facility. The powder samples (mass ~20g) were wrapped in a thin Al-foil and mounted inside a thin-walled cylindrical Al-can, which was cooled down to 4.5 K inside a closed-cycle refrigerator with He-exchange gas around the samples. The measurements were performed with various selected incident neutron energies (E$_i$) between 20 meV and 100 meV. Inelastic neutron scattering investigations are ideal for the present study as they give direct information of spin-spin correlations as well as to as well as to enable a direct estimation of spin gap value, its wavevector and temperature dependent behaviour. High resolution neutron powder diffraction measurements on CeOs$_2$Al$_{10}$ sample were carried out between 4 K and 80 K using the HRPD diffractometer at ISIS facility.

## III. Results and discussions

The CeT$_2$Al$_{10}$ (T=Fe, Ru and Os) compounds crystallize in the formation of a caged orthorhombic YbFe$_2$Al$_{10}$-type structure (space group Cmcm, No. 63) (Fig. 3 bottom). In this caged-type structure the Ce atom is surrounded by a polyhedron formed by 4 T (=Ru/Fe/Os) and 16 Al atoms and forms a zigzag chain along the orthorhombic *c*-axis [78]. The lattice parameters, unit cell volume and the selected Ce-Ce, Ce-T and Ce-Al interatomic distances are given in Table II. One can see from Table-II that the lattice parameters (*a, b, c*) are nearly similar for Os and Ru, but decrease noticeably when going from Os and Ru to the member comprised of the smaller Fe. The lattice parameters *a* and *c* contract more than *b* when going from Os or Ru to Fe. This observation is in agreement with the charge density distribution study on CeT$_2$Al$_{10}$ (T=Ru, Fe) and the crystal structure parameters of RT$_2$Al$_{10}$ [79-80]. This study shows that the lattice parameters of *a-, b-,* and *c*-axes exhibit an anisotropic contraction when Ru is replaced by Fe in RT$_2$Al$_{10}$, in contrast to the isotropic contraction simply expected from the smaller ionic radius of Fe compared to Ru. This anisotropic contraction of the YbFe$_2$Al$_{10}$-type crystal structure originates in the deviation from linearity of the zigzag chain formed by T and Al bond along the *a-* and *c*-axes that is larger than that along the *b*-axis [79]. The anisotropic contraction of the lattice parameters are expected to be the origin of the anisotropic c–f hybridization in CeT$_2$Al$_{10}$. The lattice parameters (especially *a-* and *c*-axes) of CeT$_2$Al$_{10}$ (T= Ru, Fe) exhibit the anisotropic deviation from the lanthanide contraction of RT$_2$Al$_{10}$ series [80]. This deviation is largest in the *a*-axis and is very small in the *b*-axis. Both the characteristic YbFe$_2$Al$_{10}$-type crystal structure and the anisotropic deviation towards the intermediate valence (for T=Fe) indicate that the largest c–f hybridization is along the *a*-axis and plays a dominant role which is associated with the unusual antiferromagnetic order in CeT$_2$Al$_{10}$ (T= Ru, Os).

The high-energy synchrotron x-ray powder diffraction study has been reported on LaRu$_2$Al$_{10}$ [79]. The data have been analysed using the maximum entropy method/Rietveld technique, which shows that the charge density between Ru and surrounding Al atoms is large, but that between La and surrounding Al atoms is small. It has been proposed that the Ru-Al10 polyhedron is the fundamental component of the crystal and the two-dimensional layer is constructed by these polyhedra in the ac-plane and is stacked along the *b*-axis by way of the Al5 atom [79].

The unit-cell volume is 861.7435Å$^3$ for T=Os, 863.635 Å$^3$ for T=Ru and 839.203 Å$^3$ for T=Fe. The small unit cell volume of T=Fe is due to the small ionic radius of Fe (3d) atoms as well as mixed valence nature of the Ce ions in this compound. An unstable valence such as this is the consequence of very strong c-f hybridization that may cause a partial transformation of the 4f electron of Ce to the conduction band. This has been also supported through the plot of lattice parameters of rare earth series, RRu$_2$Al$_{10}$ and RFe$_2$Al$_{10}$ [80]. The change in the unit cell volume is about ~3% while going from Ru to Fe. Furthermore, the nearest neighbour Ce-Ce and Ce-T distances are nearly the same for T=Ru and Os, but smaller for T=Fe. It is unusual to have high magnetic ordering temperatures in T=Ru and Os compunds as they have a Ce-Ce separation distances as large as ~5.2 Å. This is greater than 3.25-3.4 Å, the Hill limit beyond which direct 4f-4f interaction should cease [81]. Therefore, direct f-f interactions can be ruled out as being responsible for the high magnetic ordering temperature in these compounds. It is interesting to note that for $CeRh_3B_2$ the Ce–Ce distance is 3:09 Å (along c-axis), which is shorter than the Hill limit, and hence large 4f–4f direct mixing is expected in this direction. These results show that the unusually high ordering temperature of $CeT_2Al_{10}$ (T=Ru and Os) probably has a different origin than that of $CeRh_3B_2$.

We have recently carried out temperature dependent lattice parameters measurements on $CeOs_2Al_{10}$ using the HRPD diffractometer at ISIS. Our study shows that the lattice parameters $a$ and $c$ decrease monotonically, while the $b$ lattice parameter reveals small change in the slope near 30 K (Fig. 5). When compared to the normalised lattice parameters (normalised to 300 K), the lattice parameter $b$ showed the weakest temperature dependent compared to the $a$ and $c$ lattice parameters. Both the lattice parameters $a$ and $c$ revealed similar temperature dependence. The different behaviour of the $b$-axis lattice parameter is thought to be connected to the degree of hybridization, as the $b$-axis exhibits the smallest value of the magnetic susceptibility [60]. Considering temperature dependent atomic distances, it was noticed that Ce-Ce and Ce-$Os_2$ distances reveal a slope change near 45 K (Fig.5), which may be related to the opening of the spin gap or hybridization gap and which has been proposed from the magnetic susceptibility and optical conductivity measurements [66-68]. We also examined the temperature dependence of the Ce-Al (all 5-distances) and Ru-Al distances, but did not find any noticeable anomaly.

## A. Physical Properties of $CeT_2Al_{10}$ (T=Fe, Ru and Os)

The scope and diversity of reported studies into the magnetic and transport properties of $CeT_2Al_{10}$ (T=Fe, Ru and Os) in polycrystalline form as well as in the single crystalline form at ambient pressure as well as in applied pressure including study at high magnetic fields [59-65] attest to the wide-spread interest in this class of compounds and the significance of their unusual behaviour. Here we confine ourselves to a discussion of the single crystal magnetic susceptibility and resistivity measurements at ambient pressure on $CeT_2Al_{10}$ (Fig. 6) [60-62]. The magnetic susceptibility of all three compounds exhibits considerable anisotropy with $\chi_a > \chi_c > \chi_b$. Of first importance is the absolute values of the overall susceptibility for all three directions, which is largest for T=Ru and decreases slightly for T=Os. On the other hand the susceptibility of T=Fe is almost factor of 3 (for $a$-axis) smaller than that of T=Ru. This marked difference in the susceptibility between T=Ru and T=Fe, indicates that the origin of the susceptibility behaviours (especially anisotropy) is different in these compounds (i.e. in $CeRu_2Al_{10}$ and $CeFe_2Al_{10}$). The magnetic susceptibility of $CeFe_2Al_{10}$ exhibits a peak at 75 K for $\chi_a$, a broad maximum at 105 K for $\chi_b$ and possibility of a broad maximum above 300 K for $\chi_c$. As Ce ions are in the valence fluctuating state as evident from $L_3$-absorption measurements [82], it is surprising to find such pronounced anisotropy in the magnetic susceptibility: valence fluctuation (VF) phenomenon is a single ion property and one would generally expect isotropic properties or very weak anisotropy [83]. Thus the observed strong anisotropy in $CeFe_2Al_{10}$ has been attributed to strong anisotropic hybridization between 4f-electron and conduction electrons [61]. Further from the peak position, the Kondo scale of on-site interaction, as measured through the Kondo temperature $T_K=3*T_{max}\chi$, is similarly anisotropic along the three directions [84]. Although the susceptibility data of $CeFe_2Al_{10}$ have been analysed on the basis of CEF model, this approach is found to show poor agreement between the data and calculated susceptibility [85], which suggests that one need is to include anisotropic hybridization effect in the CEF calculations. It is interesting to compare this situation with the tetragonal intermediate valence (IV) or valence fluctuating system $YbB_4$ which exhibits a similarly strong anisotropy in the physical properties [86]. The *c*-axis susceptibility of $YbB_4$ follows Curie-Weiss behavior with smaller crystalline electric field splitting, but the *ab*-plane susceptibility shows a typical temperature dependence as observed in IV or VF compounds [87]. This implies that $YbB_4$ and $CeFe_2Al_{10}$ are rare examples of anisotropic behaviour in IV/VF state. In general the anisotropic susceptibility behaviour has not been noticed in cubic Kondo semiconductors such as $Ce_3Pt_3Bi_4$, $SmB_6$, $U_3Ni_3Sb_4$ and $YbB_{12}$, and hence DC-susceptibility measurements along the cubic-axes do not provide any information on the anisotropic hybridization. It is interesting to note that inelastic neutron scattering investigation on the cubic VF system $CePd_3$ has provided direct evidence on the ***k***-dependent, hybridization [88]. In an orthorhombic Kondo semimetal CeNiSn, the anisotropic gap has also been observed in the inelastic neutron scattering study [89], which has been explained theoretically by considering ***k***-dependence of the hybridization matrix element between *f*- and conduction electrons that can give rise to an anisotropic hybridization gap of heavy fermions if the filling of electrons corresponds to that of the band insulator [90-91]. These theories show that the most interesting case occurs when the hybridization vanishes along some symmetry axis of the crystal reflecting a particular symmetry of the crystal field and hence the wave functions of the CEF ground state are very important ingredients [90-91].

$\chi_a(T)$ of $CeOs_2Al_{10}$ exhibits a broad peak near 45 K, which is well above the ordering temperature $T_N$=28.5 K (Fig.6). This peak is associated with an opening of the spin gap well above $T_N$, which has been supported through the optical conductivity study [65-68] and our inelastic neutron scattering study [92] also discussed in Section-C below. A very similar behaviour may exist for $\chi_a(T)$ for $CeRu_2Al_{10}$ compound, however in this case $T_N$ and this susceptibility peak are at similar temperatures and therefore masks the effect (Fig.6). It is agreed that Ce ions in $CeRu_2Al_{10}$ are in $Ce^{3+}$ ionic state ($4f^1$), but $CeOs_2Al_{10}$ shows a strong hybridization effect with valence close to $Ce^{3+}$ [60]. This has been supported through the Ce $L_3$-edge measurements near room temperature [82]. For the $CeOs_2Al_{10}$ and $CeRu_2Al_{10}$ compounds the magnetic moment in the ordered state is aligned along the *c*-axis (and not along *a*-axis, which is an easy axis of the magnetization in the paramagnetic state and supported from the CEF analysis [85, 93]) with $\mu_{ord}$ = 0.42 and 0.29 $\mu_B$/Ce for T = Ru and Os, respectively [94, 95]. Strigari et al. have shown that the crystal electric field (CEF) scheme obtained from soft X-ray absorption spectroscopy can explain the easy axis and the small magnetic moment on $CeRu_2Al_{10}$ and $CeOs_2Al_{10}$ [85, 93], which are in agreement with our inelastic neutron scattering data [92]. Kunimori et al. also performed mean-field calculations and showed that the CEF scheme by itself is not enough to account for the magnetic order below $T_N$ and pointed out the importance of the conduction- and f-electron (c-f) hybridization for the unusual magnetic ordered state [96]. Kondo et al. speculated that the strong c-f hybridization suppresses the spin degrees of freedom along the *a*-axis and forces the ordered moment $\mu_{ord}$ to be along the *c*-axis in place of the *a*-axis [97]. By applying a magnetic field along the *c*-axis, a spin-flop transition is observed for $\mu_{ord}$ from *c*-axis to *b*-axis, even though the *a*-axis susceptibility is the highest in the paramagnetic state, when the magnetic field is beyond a characteristic value of about 4 T for T=Ru, [97]. In spite of the unusually high ordering temperatures, it has been demonstrated that the magnetic ground state of $CeRu_2Al_{10}$ and $CeOs_2Al_{10}$ is very unstable and changes dramatically with very small electronic perturbation [98, 99]. It has been shown that the direction of the ordered state magnetic moment is along *a*-axis (as expected, based upon on CEF analysis) for very small Ir and Rh doped systems i.e. $Ce(Os_{1-x}Ir_x)_2Al_{10}$ (x=0.08) [98] and $Ce(Ru_{1-x}Rh_x)_2Al_{10}$ (x=0.03-0.1) [99].

The temperature dependent resistivity of $CeT_2Al_{10}$ (T=Fe, Ru and Os) single crystals exhibits strong anisotropy that is larger at low temperature for T=Fe (see Fig.6). The temperature dependent resistivity of T=Fe passes through a broad peak near 50 K for current along the *a*-axis and near 100 K for *c*-axis, which is due to an onset of coherence among the 4f-electrons which scatter conduction electrons. No clear distinct coherence anomaly was observed in the high temperature for T=Ru and Os, which is due to the proximity of coherence and magnetic ordering in these two compounds. Furthermore the resistivity of $CeFe_2Al_{10}$ exhibits Kondo semiconducting behavior with an activation energy of 15 K, while NMR and heat capacity studies reveal a larger value of the gap, 125 K and 100 K, respectively [61, 69, 70]. The Kondo semiconductor behavior observed in $CeFe_2Al_{10}$ bears similarity with that of the well-known Kondo semiconductors CeNiSn and CeRhSb [100, 101]. Although the ground state of $CeRu_2Al_{10}$ is metallic (Fig.6), its resistivity increases abruptly at $T_N$ and exhibits a maximum at 23 K and then decreases with temperature, but with a visible inflection point at $T_N$. On the other hand the resistivity of T=Os exhibits semiconducting or semimetallic behaviour even below $T_N$, which is opposite to that observed in T=Ru, and rather qualitatively similar to that observed in T=Fe. Furthermore, a pressure study has shown that the resistivity of T=Ru at low temperatures increases and the ground state became semiconducting at 2 GPa [102]. The increase in the resistivity of T=Ru becomes suppressed above 3 GPa and the ground state again becomes metallic above 5 GPa. The pressure range of semimetallic behaviour in T=Ru is probably connected with proximity of the sharp 4f-electron resonance to the Fermi energy, admixed with effects of partial gapping in the conduction band. The magnetic contribution to the resistivity of T=Ru above 4 GPa shows a maximum (at 100 K), which seems to be attributed to Kondo coherence. The systematic changes in the temperature dependent resistivity were also observed in applied pressure for T=Fe and Os. At an ambient pressure $CeOs_2Al_{10}$ has a Kondo semimetallic ground state [102], but with increasing pressure at 2 GPa and above its ground state becomes metallic, and also exhibits a maximum in the resistivity near 100 K, which is opposite to that observed in T=Ru. Further the ground state of $CeFe_2Al_{10}$ changes to metallic at a very small pressure of 0.8GPa. Interestingly, the long-range magnetic ordering suddenly disappears under pressures at 2 GPa for T = Os and at 4 GPa for T = Ru [102]. The phase transition at $T_N$ exhibits maximum ($T_N$=33 K) at 2 GPa for T=Ru, while for T=Os it does not change much with pressure up to 1.5 GPa and the suddenly decreases to zero at 2 GPa [102]. This behaviour of T=Ru and Os allows us to position them accordingly on the Doniach phase diagram, see Fig.2.

In order to investigate a structural change induced by pressure, where the magnetic order disappears, synchrotron X-ray diffraction study at room temperature under high pressure up to 10 GPa for T=Fe, Ru and Os has been reported [103]. This study shows that there is no detectable structural phase transition up to 10 GPa at 300K. The lattice parameters *a, b, c* decrease linearly with pressure for all three compounds. This suggests that the pressure induced nonmagnetic ground state in $CeRu_2Al_{10}$ and $CeOs_2Al_{10}$ is not due to any structural phase transition [103]

## B. μSR measurements

μSR is an ideal local probe to investigate small moment magnetism in strongly correlated electron systems, for a review see refs. [105, 106]. In fact, μSR technique is so sensitive to small magnetism that the distribution of fields arising from nuclear moments is easily investigated. The high sensitivity of μSR technique, compared to neutron diffraction, is due to the large gyromagnetic ratio ($\gamma_\mu$ =851.6 Mrad/s/T) of the muon. From μSR study the distribution and dynamics of the internal field at muon sites can be probed. Due to being a local probe, μSR study provides information on volume fraction of different phases as well as different spin dynamic information for multi sites systems. There are many examples in heavy fermion systems where the ordered state moment is very small, due to screen of the local moment (f or d-moment) by conduction electrons through the Kondo coupling [105]. Further, μSR technique has been also used to investigate spin gap formation in many strongly correlated electron systems [107] as well as quantum criticality and NFL-behaviour in heavy fermion systems [105]. In this section we will discuss the μSR measurements on selected compounds of interest among $Ce(Ru_{1-x}Fe_x)_2Al_{10}$ (x=0 to 1) and $CeOs_2Al_{10}$, which exhibit very small moment ordering and also on spin gap formation in non-ordered ground state (x=0.8 to 1). For comparison we will also present results on stable 4f-moment systems, $NdT_2Al_{10}$ (T=Fe and Os) as well as the 3d-moment system $YFe_2Al_{10}$.

# 1. μSR study on Ce(Ru$_{1-x}$Fe$_x$)$_2$Al$_{10}$ x=0 to 1

A detailed μSR investigations on Ce(Ru$_{1-x}$Fe$_x$)$_2$Al$_{10}$ and CeOs$_2$Al$_{10}$ are reported in refs. [82, 92, 94, 108]. Figure 7 (a-h) shows the zero-field (ZF) μSR spectra at various temperatures of Ce(Ru$_{1-x}$Fe$_x$)$_2$Al$_{10}$ (x=0, 0.3, 0.5, 0.8 and 1) from refs. [82]. At 30-35 K a strong damping at shorter time (Fig.7d-h) and the recovery at longer times have been observed, which is a typical muon response to nuclear moments, described by the Kubo-Toyabe (KT) formalism [109], arising from a static distribution of the nuclear dipole moments. Above the anomaly at 28 K, i.e. in the paramagnetic state, the μSR spectra have been analysed by the following equation (see Figs. 7d-h):

$$G_z(t) = A_0 \left( \frac{1}{3} + \frac{2}{3}(1-(\sigma_{KT}t)^2)\exp\left(-\frac{(\sigma_{KT}t)^2}{2}\right) \right) \exp(-\lambda t) + BG \qquad (2)$$

where A$_0$ is the initial asymmetry, σ$_{KT}$ is nuclear depolarization rate, σ$_{KT}$/γ$_μ$ =Δ is the local Gaussian field distribution width, γ$_μ$ is the gyromagnetic ratio of the muon, λ is the electronic relaxation rate and BG is a constant background. The value of σ$_{KT}$ was found to be 0.32-0.36 μs$^{-1}$ (depending on x) from fitting the spectra of 35/30 K to Eq.(1) and was found to be temperature independent above 35 K. It is to be noted that using a similar value of the σ$_{KT}$ Kambe et al [108] have suggested 4$a$ as the muon stopping site in CeRu$_2$Al$_{10}$, while for CeOs$_2$Al$_{10}$ the muon stopping site was assigned to the (0.5, 0, 0.25) position [92]. Recently Guo et al [99] have investigated Ce(Ru$_{1-x}$Rh$_x$)$_2$Al$_{10}$ (x=0.05 and 1) using μSR and their dipolar fields calculation supports the (0.5, 0, 0.25) site for muons.

Interestingly the μSR spectra of Ce(Ru$_{1-x}$Fe$_x$)$_2$Al$_{10}$ for x=0 to 0.5 below 27 K reveal coherent frequency oscillations and they have been described by two oscillatory terms and an exponential decay, as given by the following equation

$$G_z(t) = \left( \sum_{i=1}^{2} A_i cos(\omega_i t + \varphi) \exp\left(-\frac{(\sigma_i t)^2}{2}\right) \right) + A_3 \exp(-\lambda t) + BG \qquad (3)$$

where ω$_i$=γ$_μ$ H$^i_{int}$ are the muon precession frequencies (H$^i_{int}$ is the internal field at the muon site), σ$_i$ is the muon depolarization rate (arising from the distribution of the internal field) and φ is the phase. Although, one might expect that there are two exponential decays each corresponding to a precession frequency. In our data, we only observed one, indicating that the two values must be very close to each other.

Fig. 8 (a-c) from ref. [82] shows the plot of internal fields (or muon precession frequencies) at the muon sites as a function of temperature for x=0, 0.3 and 0.5. This shows that the internal fields appear just below 27 K for x=0, below 26 K in x=0.3 and below 22 K in x=0.5.

Given the amplitudes of the oscillations this shows clear evidence for long-range magnetic ordering of the Ce moments. Further it is very important to see that the asymmetry $A_3$ drops nearly 2/3 and the relaxation rate exhibits small drops at $T_N$ for x=0, 0.3 and 0.5 (see inset Fig.8c left bottom), which confirms that the magnetic ordering is observed in the full volume of the samples and hence is bulk in nature. This is also in agreement with the phase diagram, x vs $T_N$ proposed from the magnetic susceptibility and heat capacity measurements as shown in Fig. 9. Further the decrease in the high internal field values with x may indicate that the ordered state Ce moment is reducing with x in agreement with the observed susceptibility behaviour [110]. The small value of the internal fields observed in x=0 to 0.5 are in agreement with the small ordered state magnetic moment of the $Ce^{3+}$ ion observed through the neutron diffraction for x=0 and x=0.5 [ 82, 94, 95].

The occurrence of the dip in x=0, which has also been observed in the µSR study of $CeOs_2Al_{10}$ (discussed below) at 10-15 K [92], may have some relation with a super lattice formation observed in the recent electron diffraction study of $CeOs_2Al_{10}$ [60]. In principle this could originate from various phenomena related to a change in the distribution of internal fields associated with a small change in the moment values or modulation. Further, for x=0.3 and 0.5 the observed anisotropy of the depolarization rates (observed in x=0) becomes smaller with increasing x. It is to be noted that although the µSR measurements on unaligned single crystals of $CeRu_2Al_{10}$ are more or less in agreement with the present study [99], however, there are small differences in the temperatures dependent behaviour of the relaxation rate σ (Fig. 8d), which require further investigations on well aligned high quality single crystal using µSR and neutron diffraction studies. Further it is to be noted that the observed two frequencies/internal fields in the present study is attributed to the possibility of two muon sites, while in the single crystal study the low frequency/internal field has been attributed to the Fermi contact field from the polarized electrons at the muon site [99].

## 2. µSR study on $CeOs_2Al_{10}$

A detailed µSR study of $CeOs_2Al_{10}$ has been reported in ref. [92], which also revealed the presence of two frequencies oscillations (or internal fields) below 29 K (Figs. 10 and 11) as seen in $CeRu_2Al_{10}$. The presence of internal field at the muon stopping site in zero-field indicates unambiguously long-range magnetic ordering of the $Ce^{3+}$ moments, which was later confirmed through the neutron diffraction investigation on the single crystal of $CeOs_2Al_{10}$ [95]. The magnetic structure proposed from the single crystal study of $CeOs_2Al_{10}$ is with propagation vector k=[0 1 0], which is equivalent to [1 0 0] for the orthorhombic Cmcm structure, with the moment of 0.3 $\mu_B$ along the c-axis [94, 95]. The c-axis moment direction is not expected based on the single ion CEF anisotropy [85], which indicates that the moment direction is governed by anisotropic two ions exchange interactions. The support of anisotropic two ions exchange interactions also comes from the spin wave measurements on the single crystal of CeOs$_2$Al$_{10}$ [111], which reveals that nearest neighbour exchange parameter is highly anisotropic and strongest along the c-axis [111, 112]. The maximum internal field observed is 50 G at 5 K in CeOs$_2$Al$_{10}$ which is smaller than 120 G at 15 K for CeRu$_2$Al$_{10}$.

The temperature dependence of the internal fields of CeOs$_2$Al$_{10}$ (Fig.11), show that there is a dip in the internal field (see Fig. 11 top left), which occurs around 15 K. The occurrence of the dip coincides with both a structural distortion observed in the recent electron diffraction study and with the onset of semiconducting behaviour in the resistivity below 15 K [61]. Moreover, below 15 K the first and the second component of the depolarisation rates also increase (Fig. 11 right). In principle this could originate from various phenomena related to a change in distribution of internal fields, but a structural transition is a likely candidate in view of the structural instability reported on this system [62]. It is to be noted that the third depolarization rate associated with zero-frequency also exhibits dramatic changes with temperature (see the inset Fig. 10 bottom-right).

## 3. µSR study on NdFe$_2$Al$_{10}$

It is interesting to note that among CeT$_2$Al$_{10}$ (T=Fe, Ru and Os), the compound with T=Fe does not order magnetically down to 40 mK, while T=Ru and Os show ordering with very small ordered state moments 0.4 and 0.3µ$_B$, respectively. Hence it is of interest to study magnetic properties of other members of the RT$_2$Al$_{10}$ series, where 4f-moment is very localised compared with the Ce-compounds. The magnetic properties of RFe$_2$Al$_{10}$ compounds with R=Pr, Nd, Sm, Tb, Dy, Er, Ho and Yb have been investigated [113]. It has been found that the magnetic ordering transitions T$_{N1}$=3.9 K (T$_{N2}$=1.5 K) for R=Nd, 16.5 K for T=Tb and 7.5 for R=Dy are very low despite a large, stable magnetic moment on the rare earth atoms (i.e 10.63µ$_B$ for Dy), while the compound with T=Ho does not order magnetically down to 1.5 K even though the magnetic moment of Ho is 10.60 µ$_B$. This indicates the important of crystal field ground state on the magnetic ordering of RFe$_2$Al$_{10}$ compounds. On the other hand YbFe$_2$Al$_{l0}$ shows a mixed valence behaviour as in CeFe$_2$Al$_{10}$ and SmFe$_2$Al$_{10}$ is Van Vleck paramagnetic [114]. Here we have carried out µSR measurements on NdFe$_2$Al$_{10}$ to investigate the observed two magnetic phase transitions in the heat capacity [96]. It is to be noted that heat capacity of NdFe$_2$Al$_{10}$ exhibits very sharp jump at T$_N$ and very broad peak around 1.8 K.

Our ZF µSR measurements of NdFe$_2$Al$_{10}$ reveal interesting behaviour with temperature (see Fig.12). At high temperature above (T$_N$) µSR spectra reveal KT-type behaviour, but below 3.8 K a clear sign of two frequency oscillations have been seen down to 1.2 K. However, at 1.2 K, frequency oscillations are visible only below 2µs time, which is most probably due to a larger values of internal fields, with broad distribution of the internal fields (which gives stronger damping) at the muon sites. We have used maximum entropy method to determine the internal fields and number of frequencies at 1.2 K and then µSR data of $NdFe_2Al_{10}$, were analysed in real time as we did for $CeT_2Al_{10}$ (T=Ru and Os).

Above the phase transition at $T_N$ and in the high temperature range the data for $NdFe_2Al_{10}$ were fitted with using KT function times the exponential decay and a constant BG term. We were able to fit the data very well with temperature independent BG term. Further the $\sigma_{KT}$ and asymmetry parameters were also temperature independent with values 0.33 ($\mu S^{-1}$), indicating very similar muon sites as in $CeT_2Al_{10}$ as discussed above.

The µSR data at 1.2 K were fitted using two frequency terms with a Lorentzian envelope (i.e. exp(-λt)), rather than a Gaussian envelope used for $CeT_2Al_{10}$, and keeping the same BG as it was used in the high temperature range above $T_N$. From the analysis it was obvious that there are two frequencies present in the data at 1.2 K (Fig. 12). The value of the higher internal field ($B_1$) is much higher than that observed in the $CeT_2Al_{10}$ compounds, which is expected as the magnetic moment of Nd ion is larger than that of the Ce ion as well as 4f-electrons are localised on the former than the latter. On the other hand the value of lower internal field ($B_2$) is very close to the highest internal field observed in $CeRu_2Al_{10}$. The temperatures dependent internal fields/frequencies of $NdFe_2Al_{10}$ are shown in Fig. 13, which reveals sharp rise in the high field/frequency (first component) just below $T_{N1}$ and saturation below 1.8 K. The lower field/frequency (second component) also increases with decreasing temperature below $T_{N1}$ and exhibits weak anomaly near 2.2 K. The µSR data are consistent with the heat capacity with a very sharp jump at $T_N$ and very broad peak around 1.8K [96].

**4. µSR study on $NdOs_2Al_{10}$**

The magnetic susceptibility, electrical resistivity, and specific heat of $ROs_2Al_{10}$ (R=Pr, Nd, Sm, and Gd) compounds, which are isostructural with a Kondo semiconductor $CeOs_2Al_{10,}$ have been reported [115-117]. The compound with R=Pr does not order magnetically down to 0.4 K, whereas the compounds with R=Nd, Sm, and Gd undergo antiferromagnetic transitions at $T_{N1}$= 2.2 ($T_{N2}$ = 1.1 K), 12.5 ($T_{N2}$ = 9.1 K, $T_{N3}$ = 5.6 K), and 18 K ($T_{N2}$ = 15 K), respectively. The magnetic susceptibility of $NdOs_2Al_{10}$ exhibits Curie-Weiss behaviour with an effective magnetic moment, $\mu_{eff}$ =3.3 $\mu_B$ and Curie-Weiss temperature, $\theta_P$=-6.0 K. The small value of $\theta_P$ indicates a weak antiferromagnetic interactions between Nd moments, which is in agreement with smaller ordering temperatures.

In order to shed light on the nature of magnetic ground state in $NdOs_2Al_{10}$, and also for comparison with other compounds reported here, especially $CeOs_2Al_{10}$, we have carried out μSR measurements on $NdOs_2Al_{10}$ and the results are shown in Figs.14 and 15. As can be seen from Fig.14 that above the magnetic ordering temperature (i.e at 4.15 K), the μSR spectra reveal KT type behaviour, similarly to what is seen in other $RT_2Al_{10}$ compounds [82, 94, 99]. Below $T_N$ loss of asymmetry was observed indicating a long range magnetic ground state below 2.2 K, which is in agreement with the heat capacity and magnetic susceptibility results [115]. The absence of clear frequency oscillations below $T_{N1}$ indicates that internal fields at the muon sites are large and are outside the maximum observable field limit (about 800 gauss) of the MUSR spectrometer at ISIS. The observable maximum field is limited due to finite pulsed width of muon pulse at ISIS. The data were fitted with KT*exp(-λ*t)+BG and the BG was estimated from 4.2 K and was kept fixed. The data fit very well to this function (see Fig. 14). Interestingly below 2.24 K the muon initial asymmetry drops to 2/3 to its high temperature value (Fig.15), which indicates that bulk nature of long range magnetic ordering of Nd-moments in $NdOs_2Al_{10}$. The electronic relaxation rate (λ) rate increases below 2.24 K (i.e at $T_{N1}$) and exhibits a peak near 2 K, while Gaussian KT relaxation rate ($\sigma_{KT}$) remains temperature independent and drops to zero below 2.24 K, which we expect as in the presence of internal field KT term becomes unity. This is due to the fact that nuclear magnetism is a factor of $10^3$ smaller than that of electronic magnetism and hence muons respond predominantly to the electronic magnetism. There is a possibility to see second phase transition at $T_{N2}$ = 1.1 K in the λ(T), however due to limitation on the minimum sample temperature during this experiment, 1.2 K, we could not investigate the second magnetic transition in our μSR study.

## 5. μSR investigations of a quantum critical behaviour in $YFe_2Al_{10}$

There have been extensive investigation on NFL and QCP behaviour on compounds having 4f or 5f-local moment using μSR measurements, for example $CeCu_{6-x}Au_x$ [118], $UCu_{5-x}Pd_x$ [119], $Y_{1-x}U_xPd_3$ [120], $CeRh_{0.85}Pd_{0.15}$ [121], $CeInPt_4$ [122] and may more for review see refs. [105]. The μSR relaxation rate exhibits power law or logarithmic divergence at low temperatures in a manner similar to what is exhibited in χ(T). The NFL and QCP behaviour in these systems have been explained using various theoretical scenarios [26]: local quantum critical point, in which Kondo temperature breaks down ($T_K$=0) at QCP [26, 123], spin density wave (SDW) model [26,46], in which $T_K$ remains finite for either sides of QCP, and valence induced QCP, in which one observes line of quantum criticality instead of a quantum critical point [124].

There are only a few examples among d-electron systems showing QCP and NFL behaviour, for example ferromagnetic quantum criticality in $Ni_{1-x}Pd_x$ [125], V-doped Cr [126] pressured induced NFL in ferromagnetic $ZrZn_2$ [127], and NFL behaviour near AFM instability $CaRuO_3$ [128]. A subclass with exotic d-electron magnetism can be formed by those compounds in which the rare-earth (or actinide) atom bears no magnetic moment ($4f^0$, La, Lu, Y and tetravalent Ce and divalent Yb). The compounds belonging to this subclass have received much attention since they offer a good opportunity to study the origin and the nature of the d-electron magnetism to compare with isostructural f-electron systems. In these systems the d-atoms do not occupy the random crystallographic sites, but are fixed in well-defined crystallographic positions. Rather conclusive information on the d-electron magnetism can therefore be obtained by means of advanced techniques such as nuclear magnetic resonance, μSR, Mössbauer effect and diffraction with polarized neutrons

Recently $YFe_2Al_{10}$ has been extensively investigated using magnetization, specific heat, and NMR over a wide range of temperature and magnetic field and zero field NQR measurements [129]. Magnetic susceptibility, specific heat, and spin-lattice relaxation rate divided by $T$ ($1/T_1T$) follow a weak power law ($\sim T^{-\alpha}$ with α=0.35 to 0.5) temperature dependence below 4 K, which is a signature of the critical fluctuations of the Fe moments (see Fig. 16). The value of the Sommerfeld-Wilson ratio and the linear relation between $1/T_1T$ and $\chi$ suggest the existence of ferromagnetic correlations in this system. The absence of a long range magnetic ordering down to 50 mK was confirmed through various studies [129]. Further the unusual $T$ and $H$ scaling of the bulk and NMR data are associated with a magnetic instability which drives the system to quantum criticality.

The χ(T) of $YFe_2Al_{10}$ obeys a Curie-Weiss behavior and the effective moment ($\mu_{eff}$ = 0.52$\mu_B$/f.u.) is rather small. The field induced moment at 6T (at 2K) $\mu_{6T} \sim 0.02\mu_B$/f.u. [129] leads to $\mu_{eff}/\mu_{6T} \approx 26$, which classifies $YFe_2Al_{10}$ as being a weak itinerant FM in the Rhodes Wohlfarth plot [129]. However, a clue about the importance of magnetocrystalline anisotropy even in this d-electron system is found in the fact that the saturation magnetization for field perpendicular to the *b*-axis at 2K amounts to only 0.02 $\mu_B$/fu [131]. Indeed, the findings of Park et al. indicate that the susceptibility in this compound is governed largely by critical fluctuations confined to the *ac*-plane and that the *b*-axis may be considered as decoupled from the magnetic fluctuations. This arrangement of the anisotropy is, interestingly, quite similar to that of the local-moment, f-electron $CeT_2Al_{10}$ members of the series (see Section B). The enhancement of ac and dc magnetic susceptibility accompanied by $\chi(T) \sim T^{-0.5}$ divergence below 10 K in a weak magnetic fields could be due to exchange enhanced q = 0 excitations and suggests the presence of spin correlations [129]. The magnetic properties of the system are tuned by applying magnetic fields wherein ferromagnetic fluctuations are suppressed and a crossover from quantum critical to Fermi-liquid behaviour is observed [129].

To shed light on the natures of critical magnetic fluctuations in $YFe_2Al_{10}$, we have carried out ZF and longitudinal applied field (LF) μSR measurements on this compound down to 60 mK. Fig.17a and b show μSR spectra in ZF and LF of 50 G at 0.06K and 3 K (2 K for 50 G). The ZF μSR spectra show a typical KT type temperature independent behaviour, which indicates that the response seen by muons is mainly dominated by the nuclear magnetism (from Fe and Al) and very small or weak contribution from d-electronic magnetism of the Fe atoms. The latter has been supported through 50 G LF data (Fig.17b), which show nearly time independent and also temperature independent behaviour of µSR spectra. In order to gain more information on the weak magnetism we also carried out field dependent study at 0.06 K (Fig.17c) and 3 K (not shown here) up to 2500 G LF. The field response at 0.06 K and 3 K, were very similar. It can be seen from Figs. 17 and 18 that between 50 G and 2500 G the electronic relaxation rate ($\lambda$) remains nearly field independent, but the initial asymmetry show weak field dependent and increases linearly with applied field. Temperature dependent data in ZF, 50 G and 2500 G LF were analysed using KT*exp(-$\lambda$*t)+BG form of the relaxation function. The fit parameters are shown in Fig.18. It is again clear that nuclear depolarization rate $\sigma_{KT}(T)$ is temperature independent. The ZF $\lambda(T)$ exhibits very weak temperature dependent, change from 0.02 $\mu s^{-1}$ at 3 K to 0.033$\mu s^{-1}$ at 0.06 K. Further in 50 G and 2500 G LF the value of $\lambda$ drops considerably and again it is nearly temperature independent. The observed nearly temperature independent behaviour of $\lambda(T)$ of $YFe_2Al_{10}$ was unexpected because the magnetization, heat capacity and NMR/NQR measurements reveal the presence of a power law diversion due to the presence of critical fluctuations [129]. The µSR measurements on f-electron based NFL and quantum critical systems exhibit very strong temperature and field dependent behaviour of $\lambda$ [105, 121]. For example FM QCP and NFL 4f-electron system $CePd_{0.15}Rh_{0.85}$, $\lambda(T)$ exhibits a power-law behavior, $\lambda(T) \sim T^{-n}$ with n~0.8, while the field dependence at 0.1 K reveals a time-field scaling of the muon relaxation function, $P_z(t, H)=P_z(t/H^\gamma)$ with $\gamma=1.0\pm0.1$ [121]. Furthermore, the exponent derived from the ZF-µSR data agrees well with the power-law behavior of the temperature-dependent susceptibility, $\chi T) \sim T^{-\alpha}$ ($\alpha=0.6$), the E/T scaling of the neutron dynamical susceptibility, as well as the magnetization-field-temperature scaling $\gamma_m=0.8\pm0.1$.

In order to check whether we can observe magnetic fluctuations in $YFe_2Al_{10}$ through neutron dynamic susceptibility, we have also carried out inelastic neutron scattering measurements using the MARI time-of-flight spectrometer between 4 K and 100 K range with incident energies between 6 and 25 meV. Our INS study did not reveal any clear sign of magnetic signal in $YFe_2Al_{10}$. This may suggest that the probing time scale (also energy scale for INS study) of the experimental techniques is very important to see critical magnetic fluctuations in $YFe_2Al_{10}$. Further µSR and high resolution low energy INS studies on a good quality single crystal of $YFe_2Al_{10}$ are highly desirable to unravel the nature of the magnetic fluctuations in this interesting d-electron system.

**C. Inelastic neutron scattering study**

Inelastic neutron scattering (INS) is the only method which allows us to determine both the spatial and time correlations of the magnetic excitations through the dipolar coupling of the magnetic moment of the neutron with the spin-correlation function of the sample [132134]. The magnetic moment of the neutron acts as a wavevector and frequency dependent magnetic field that probes the dynamic magnetic response of the sample. The energy of thermal neutrons is comparable to the energy scale of magnetic excitations and lattice vibration (or phonons) in solid state materials and hence INS is suitably disposed to provide vital information on magnetic exchanges as well as lattice dynamics. INS study has been extensively used to understand the origin of superconductivity in Cu-based and Fe-based materials as well as unconventional superconductivity in HF by investigating high- and low-energy dispersive spin excitations, spin gap formation and spin resonances and phonon dispersions [135-137]. Further INS study has been extensively used recently to investigate low dimensional quantum magnets in which the ground state forms nonmagnetic singlet state (total spin S=0 with formation of short rage AFM dimers) down to the lowest energy [138]. INS study directly probes the energy scale of singlet (S=0) to triplet (S=1) state, by flipping neutron spin from -1/2 to +1/2 (total spin change of neutron is 1), and hence provides information on the interdimer and intradimer exchange interactions [138, 139]. In addition to this the INS study has been used to investigate magnons condensation in the Bose-Einstein condensate (BEC) state of many oxides based magnets [140] as well as to investigate quantum fluctuations near QCP and NFL state in HF systems [141]. In the present systems, $CeT_2Al_{10}$ (T=Ru and Os), INS technique was used to unravel the contribution of gapped spin wave ground state, arising from single ion CEF anisotropy, and the gap formation due to hybridization between 4f-electrons and conduction electrons or spin dimer formation. The latter two contributions are also investigated in $Ce(Ru_{1-x}Fe_x)_2Al_{10}$ with x=0.8 and $CeFe_2Al_{10}$, which have nonmagnetic ground state down to 1.2 K and 40 mK, respectively [82].

**1. INS study on $Ce(Ru_{1-x}Fe_x)_2Al_{10}$ (x=0 to 1)**

To date, two groups have investigated and reported on powder samples of $CeRu_2Al_{10}$ using inelastic neutron scattering and both have confirmed the formation of a spin gap of 8 meV in the inelastic response, together with its strongly temperature dependent character [82, 142]. On the other hand, the Q-dependent response following $Ce^{3+}$ magnetic form factor in the powder sample may indicates the importance of single-ion interactions or strong intensity of magnon near the zone boundary arising from powder averaging effect [82, 142]. There is also a possibility of existence of a gap type response above $T_N$ up to 34 K in agreement with the optical conductivity study [82, 66-68]. The gap is nearly temperature independent very close to $T_N$, but then abruptly transforms into a broad quasi-elastic/inelastic response above $T_N$ [82, 142]. By raising the temperature still further (above 35-40 K), the INS response becomes very broad, with quasi-elastic character [82, 142]. The observation of a spin gap in these compounds is in good agreement with predictions based on a theoretical model for a spin-dimer formation pertinent to this class of compounds which has recently been put forward by Hanzawa [143, 143]. However, recent spin wave studies by Robert et al on single crystals of CeRu$_2$Al$_{10}$ [112] using a triple axis spectrometer revealed the gapped spin wave excitations (gap ~ 4-5 meV at AFM zone centre). Near the AFM zone centre there are two spin waves modes, which are expected considering that there are two spin per primitive unit cell. The zone boundary energy is about 8 meV in all directions. These observations are in full agreement with our spin waves investigation on CeRu$_2$Al$_{10}$ single crystal using a time-off-flight neutron scattering technique on the MERLIN spectrometer [111]. The analysis of spin waves by Robert et al and also by us indicates that the nearest neighbour (NN) interactions are highly anisotropic, i.e. exchange along the *c*-axis is the strongest [112]. This result supports the observation of the magnetic structure with magnetic moment along the *c*-axis and not along the *a*-axis as expected from the CEF anisotropy in the paramagnetic state [93]. Even though we have a clear sign that the INS at 4.5 K is due to spin wave in CeRu$_2$Al$_{10}$, it is still an open question to find direct evidence of INS scattering arising from the hybridization gap or dimer formation. In order to unravel these two contributions it was necessary to investigate Ce(Ru$_{1-x}$Fe$_x$)$_2$Al$_{10}$ alloys with x=0.8 and 1 where magnetic ordering disappears and hence we can directly study the INS response arising from the hybridization gap formation or dimer formation. A detailed report on this has been published by some of us [82] and here we will give a brief discussion on this.

Figure 19 shows INS scattering at low-Q from Ce(Ru$_{1-x}$Fe$_x$)$_2$Al$_{10}$ with x=0, 0.3, 0.5, 0.8 and 1. There is a clear sign of magnetic excitation centred around 8 meV in x=0 and around 10-12 meV for x=0.8. The value of the peak position can be taken as a measure of the spin gap energy in these compounds [145]. It was found that the temperature dependence of the observed spin gap in CeRu$_2$Al$_{12}$ and theoretically predicted spin gap by the model of Hanzawa [143, 144] have similar behaviour just below T$_N$, but there is clear evidence in the experimental data to support the existence of the spin gap just above T$_N$ (possibly up to 33 K) in x=0 and also in x=0.3 (up to 35K) [82]. The optical study on CeRu$_2$Al$_{10}$ also shows the existence of a energy gap above T$_N$ through the effective electron number $N_{eff}$, which is related to the gap $N_{eff} \sim \Delta^2_{opt}$ [67]. A very similar situation has also been observed for CeOs$_2$Al$_{10}$ through an optical study [68], where a CDW gap (or $\Delta_{opt}$) exists up to 39 K, and also from our recent INS study [92] discussed below. If we take the value of the quasi-elastic linewidth as a measure of Kondo temperature T$_K$ (just above T$_N$, ideally one takes the value at T=0) then it shows that T$_K$ increases from 52(3) K in x=0, 83(5) K in x=0.3, to 110(10) K in x=0.5.

It is clear from the INS study that a spin gap exists at 7 K and the response becomes quasi-elastic at 94 K in Ce(Ru$_{1-x}$Fe$_x$)$_2$Al$_{10}$ with x=0.8 and 1 [82]. A notable feature of the spin gap energies of these compounds measured through INS is their universal scaling relationship with the Kondo energy (T$_K$) derived from the maximum in the susceptibility [145, 146, 147]. According to the single impurity model [145, 148], we can estimate the high temperature Kondo temperature $T_K$ through the maximum $T_{max}(\chi)$ in the bulk susceptibility as $T_K = 3*T_{max}(\chi) = 150$ K (12.92 meV) for x=0.8. This shows that the spin gap of 10 meV observed through the INS study is in agreement with the scaling behavior.

The Q-dependent intensity of the 10 meV excitation in x=0.8 exhibits a clear peak near Q = 0.8 Å$^{-1}$, which is different from that observed for many spin gap systems that do not exhibit long-range magnetic order [145]. The fit to the dimer structure factor $I(Q) \sim sin(Q\ d)/(Q\ d)$, where d is the Ce-Ce distance, gave d=5.07(4)Å, which is close to d=5.21(4)Å estimated through neutron diffraction study at 300 K [82], and which in turn is supportive of dimer formation. Another possible interpretation of the observed spin gap in x=0.8 and 1 could be an anisotropic spin gap opening only on a small part of the Fermi surface or along a specific direction in Q-space. This is somewhat similar to the spin gap observed only along [0 0 l] direction in the Kondo semimetal CeNiSn [89].

## 2. Spin gap above magnetic ordering in CeOs$_2$Al$_{10}$

On CeOs$_2$Al$_{10}$ a preliminary study of a spin gap formation using INS study has been reported by some of us [92]. Here we present a detailed analysis of temperature dependence of the spin gap formation in CeOs$_2$Al$_{10}$. Besides our previous INS study, we have also obtained additional INS data using the high flux neutron spectrometer MERLIN (see Fig. 20) and carried out detailed analysis (see Fig. 21). The phonon contribution was subtracted using the data of nonmagnetic reference compound LaOs$_2$Al$_{10}$. From Fig. 20 it is clear now that we have strong magnetic scattering centred near 11 meV with dispersion coming out from Q = 0.6-0.7Å$^{-1}$, which is the AFM zone centre [0 1 0], which indicates that a part of INS response is associated with the gapped spin waves [111, 112]. The presence of clear dispersive gapped spin wave (with gap of ~5 meV) has also been observed in our single crystals INS study of CeOs$_2$Al$_{10}$ [111]. Considering in particular the temperature dependent scattering (see Fig. 21 bottom), it is clear that magnetic scattering remains practically temperature independent up to 25 K and then decreases with increasing temperature, but remains inelastic up to 35-38 K (i.e. above $T_N$ = 28.5 K) and from 40 K and above the scattering becomes quasi-elastic. A very similar behavior has been observed in CeRu$_2$Al$_{10}$ as discussed above [82]. It is to be noted that the optical study reveals the presence of charged density wave (CDW) gap above $T_N$ in CeOs$_2$Al$_{10}$ [66, 67]. The value of the optical gap observed at 4 K is $\Delta_{opt}$ = 20 meV, which is almost double that of the spin gap (powder average gap) observed through INS study.

Figure 21 shows the temperature dependent parameters estimated from the fit to the INS data (filled circles are inelastic peak fits and open circles are quasi-elastic fits) of CeOs$_2$Al$_{10}$. A detailed account of the data analysis and fitting procedure are given in ref. [82]. Fig. 20(a) shows the estimated magnetic susceptibility from INS fit (symbols). Therefore, we assumed that van Vleck contribution from the high energy CEF is small at low temperature. The estimated susceptibility is close to that measured using a SQUID magnetometer on a powder sample of $CeOs_2Al_{10}$ shown by a green solid line in Fig. 21(a). Fig.21(b) shows the temperature dependent linewidth, $\Gamma(T)$, and Fig. 21(c) shows the temperature dependent peak position, $\Delta(T)$, (i.e. spin gap). It is clear that $\Gamma(T)$ decreases below $T_N$. We have analysed $\Gamma(T)$ using two models: (1) $\Gamma(T) \sim T^2$ (see solid line in Fig.21b) and (2) exponential behavior, $\Gamma(T) \sim \exp(-\Delta(0)/k_BT)$ (see dotted line in Fig. 21b), where $k_B$ is Boltzmann's constant. The exponential relation for $\Gamma(T)$ is found to describe the data much more reliably with gap value of $\Delta(0) \sim 10.2(1.0)$ meV, which is in good agreement with the peak position observed at 4.5 K. For comparison we have also plotted temperature dependent optical gap, $\Delta_{opt}/2$ (divided by factor 2), in Fig. 21(c) (left y-axis) [67]. A very similar temperature dependent behaviour has been observed for both INS gap and optical gap, i.e. both gaps exist up to 39 K. This may indicate either the presence of spin waves above $T_N$ or the involvement of a hybridization gap (or dimer gap) above $T_N$. Considering that gapping is also observed in the optical study above $T_N$ and also a broad peak in the DC-susceptibility near 45 K (see Fig.6), it is likely that the observed gap above $T_N$ is associated with the hybridization (or dimer) gap. Support for the existence of a gap above $T_N$ can also be found through Ce-Ce and Ce-$Os_2$ distances (see Fig. 5). If we take the value of the quasi-elastic linewidth as a measure of Kondo temperature $T_K$ (as we did for $CeRu_2Al_{10}$) then estimated $T_K$ is 92(5) K for $CeOs_2Al_{10}$, which is almost double that of $CeRu_2Al_{10}$. This indicates that the hybridization between 4f-electrons and conduction electrons is stronger in $CeOs_2Al_{10}$ than in $CeRu_2Al_{10}$.

## IV. Conclusions and future work

We have carried out a suite of comprehensive µSR and inelastic neutron scattering measurements on $Ce(Ru_{1-x}Fe_x)_2Al_{10}$ (x=0, 0.3, 0.5, 0.8 and 1), $CeOs_2Al_{10}$ and reference compounds, $NdFe_2Al_{10}$, $NdOs_2Al_{10}$ as well as on the 3d-electron quantum magnet $YFe_2Al_{10}$ to understand the unusual magnetic phase transition and spin gap formation. Confirming the long-range antiferromagnetic nature of the phase transition observed in the two compounds $CeRu_2Al_{10}$ and $CeOs_2Al_{10}$ using these techniques was a major milestone and a step forward in advancing our understanding of correlated phenomena in these systems. Our µSR studies reveal the presence of two frequency oscillations below the magnetic ordering temperatures for the Ce-based compounds as well as in $NdFe_2Al_{10}$, which provides direct evidence of the long-range magnetic ordering. On the other hand the µSR spectra of $NdOs_2Al_{10}$ reveal strong damping and loss of initial asymmetry without a clear sign of frequency oscillations below $T_N$, indicating that the internal fields at the muon stopping sites are larger than 800 G (maximum field that can be investigated at ISIS facility. Further no clear sign of magnetic critical fluctuations was observed down to 60 mK in µSR study of $YFe_2Al_{10}$. For the Ce compounds exhibiting long range magnetic ordering, the temperature dependence of the µSR frequencies and the muon depolarization rates follow an unusual behavior with cooling of the sample below $T_N$, pointing to the possibility of another phase transition at low temperatures. Further, the µSR spectra of x=0.8 and 1 down to the lowest temperature confirm the paramagnetic ground state. The inelastic neutron scattering (INS) study has established the formation of a spin energy-gap with an energy scale of ~8 meV in the magnetic ordered compounds x=0, 0.3 and 0.5 as well as a spin gap of 11 meV in $CeOs_2Al_{10}$.

Our INS study reveals the possibility of an INS peak existing above $T_N$ in $CeRu_2Al_{10}$ and $CeOs_2Al_{10}$, which may indicate that the origin of the spin gap is associated with the hybridization gap or spin dimer formation. Moreover, the INS results of paramagnetic compounds x=0.8 and 1 reveal the presence of inelastic peaks (or spin gap) at 10 and 12.5 meV, respectively and at high temperature the response transforms into a quasi-elastic line. Further the investigations of very small Rh doping in $CeRu_2Al_{10}$ and Ir doping in $CeOs_2Al_{10}$ reveal that the magnetic moment orients along the *a*-axis as expect based on the single ion crystal field anisotropy [98, 99], which suggests that magnetic ground state and 4f-electrons and conduction electrons hybridization of these compounds are highly sensitive to electron doping. Further support of the presence of the anisotropic hybridization in $CeOs_2Al_{10}$ comes from a lightly hall doping (i.e 3% Re doping on Os site), which indicates a considerable reduction in the ordered state Ce moment ~0.18 $\mu_B$ with reduce $T_N$ = 20 K [ 180].

We believe the coexistence of the Kondo semiconducting state with spin-gap formation and magnetic order in $CeT_2Al_{10}$ (T=Ru, Os) to be unique among 4f-electron systems and it poses a perplexing new ground state for the strongly correlated class of materials. Considering the observation of a spin gap above $T_N$ in powder sample of $CeT_2Al_{10}$ (T=Ru and Os) and well defined spin wave at 4.5 K in the single crystal samples, it is highly desirable to investigate $CeT_2Al_{10}$ single crystals using INS above $T_N$. Further it would be very interesting to investigate spin waves, spin gap and crystal field excitations in $Ce(Ru_{1-x}Rh_x)_2Al_{10}$ and $Ce(Os_{1-x}Ir_{x)2}Al_{10}$ single crystals, where the ordered state magnetic moment is along the *a*-axis as expected base on CEF anisotropy, and anisotropic exchange and hybridization seem to have weaken, to unravel the role of hybridization on the magnetic properties of the parent compounds.

## Acknowledgement:

We acknowledge interesting discussion with Andrea Severing, Dimitry Khalyavin, Peter Baker, Stephen Cottrell, Pascale Deen, Ross Stewart, Pascale Manuel, Amir Murani, Peter Riseborough, Qimiao Si and Piers Coleman. DTA and ADH would like to thank CMPC-STFC, grant number CMPC-09108, for financial support. The work at Hiroshima University was supported by a Grant-in-Aid for Scientific Research on Innovative Area "Heavy Electrons" (20102004) of MEXT, Japan. AMS thanks the SA-NRF (Grant 78832) and UJ Research Committee for financial support. AB thanks the FRC of UJ and ISIS-STFC for funding support.

.

**Figure captions**

Fig. 1 (color online) (a) Single impurity Kondo effect, a schematic view of the conduction electron band and localised 4f-electron, and resonant density of states (DOS) showing build up of the fermionic resonance near the Fermi level. The width of the resonance peak gives an estimation of Kondo temperature, $T_K$. (b) Kondo lattice case, hybridized band picture showing renormalized bands, lower ($E_K^-$) and upper ($E_K^+$) hybridized bands, and direct gap, $\Delta_{dir}$ (we called charged gap, $\Delta_{char}$) at q = 0 and indirect gap, $\Delta_{ind}$ (we called spin gap, $\Delta_{spin}$) at q ≠ 0 and the gap in resonant density of states (DOS). (c) The *4f*-weight factors of the upper ($E_K^+$) and lower ($E_K^-$) bands as a function of wave vector, taken from Riseborough [18].

Fig.2 (color online) Generic heavy fermion phase diagram (originated from Doniach's model [47]): coupling strength (δ) versus the characteristic temperature (from ref. [149]). The proposed position of $CeT_2Al_{10}$ (T=Fe, Ru and Os) compounds on the phase diagram are obtained from the pressure dependent data on $T_0/T_N$ [102].

Fig.3 (color online) (bottom) The caged type orthorhombic unit cell of $CeT_2Al_{10}$ (T=Fe, Ru and Os) and (top) magnetic structure of $CeRu_2Al_{10}$ from ref. [94]

Fig.4 (color online) de Gennes factor versus ordering temperature of $RRu_2Al_{10}$ (R=rare earths) (top) and $ROs_2Al_{10}$ (bottom).

Fig.5 Temperature dependence of the lattice parameters and unit cell volume (left) and selected interatomic distances of $CeOs_2Al_{10}$ investigated using the high resolution neutron powder (HRPD) diffractometer at ISIS facility.

Fig.6 (a-c) Temperature dependence of the magnetic susceptibility of $CeT_2Al_{10}$ (T=Fe, Ru and Os) and (d-f) temperature dependence of the electrical resistivity of $CeT_2Al_{10}$ (T=Fe, Ru and Os) from refs. [60-62]

Fig. 7 (color online) Zero-field (ZF) μSR spectra of $Ce(Ru_{1-x}Fe_x)_2Al_{10}$ (x=0 to 1) at various temperatures from ref [82]. The solid lines show a fit as discussed in the text.

Fig. 8 (color online) Fit parameters of zero-field (ZF) μSR spectra of $Ce(Ru_{1-x}Fe_x)_2Al_{10}$ (x=0 to 1), internal fields vs temperature (left), and depolarization rate vs temperature (right) from ref.[82]. Two distinct frequencies have been found, leading also to two depolarization rates which are plotted in red and black symbols. The inset in (c) shows the temperature dependence of the muon initial asymmetry for three components from Eq.(2).

Fig. 9 (color online) Phase diagram of $Ce(Ru_{1-x}Fe_x)_2Al_{10}$ [from Nishioka et al [46]] and $Ce(Os_{1-x}Fe_x)_2Al_{10}$. The phase transition disappears sharply at $x_C$ = 0.8 in $Ce(Ru_{1-x}Fe_x)_2Al_{10}$ and at $x_C$ = 0.6 in $Ce(Os_{1-x}Fe_x)_2Al_{10}$. The jump in the heat capacity, $\Delta(C/T)$, at $T_N$ decreases gradually with increasing x. Note that the phase diagram of $Ce(Os_{1-x}Fe_x)_2Al_{10}$ overlaps that of $Ce(Ru_{1-x}Fe_x)_2Al_{10}$ when the *x* value is shifted by 0.25.

Fig. 10 (color online) Zero-field (ZF) μSR spectra of $CeOs_2Al_{10}$ at various temperatures from ref. [92]. The solid lines show the fit (see text).

Fig. 11 (color online) Fit parameters of zero-field (ZF) μSR spectra of $CeOs_2Al_{10}$, internal fields vs temperature (left), and depolarization rate vs temperature (right) from ref. [92]. Two distinct internal fields (or frequencies) have been found, leading also to two depolarization rates which are plotted in red and black symbols.

Fig. 12 (color online) Zero-field (ZF) μSR spectra of $NdFe_2Al_{10}$ at various temperatures. The solid lines show a fit as discussed in the text.

Fig. 13 (color online) Temperature dependence of the internal fields at the muons' stopping sites of $NdFe_2Al_{10}$.

Fig. 14 (color online) Zero-field (ZF) μSR spectra of $NdOs_2Al_{10}$ at various temperatures. The solid lines show the fit (see text) and the dotted line shows the background.

Fig. 15 (color online) (a) Temperature dependence of the muon initial asymmetry and, (b) KT-depolarization rate ($\sigma_{KT}$, red open circles) and electronic relaxation rate ($\lambda$, filled squares) of $NdOs_2Al_{10}$.

Fig. 16 (a) Magnetic susceptibility as function of temperature (from ref. [129]) and, (b) heat capacity divided by temperature as function of temperature of $YFe_2Al_{10}$. The solid line shows the power law ($T^{-n}$) behaviour.

Fig.17 (a) Zero-field (ZF) μSR spectra at 0.06 K and 3 K, (b) μSR spectra measured in 50 G longitudinal field (LF) at 0.06 K and 2 K and, (c) field (LF) dependence of μSR spectra at 0.06 K.

Fig. 18 (color online) Temperature dependence of the muon initial asymmetry (top) and electronic relaxation rate ($\lambda$) of $YFe_2Al_{10}$ measured in 0 G (ZF), 50 G and 2500 G and and KT-depolarization rate ($\sigma_{KT}$, left y-axis) measured in 0 G (ZF).

Fig. 19 (color online) Q-integrated inelastic scattering intensity versus energy transfer of $Ce(Ru_{1-x}Fe_x)_2Al_{10}$ for x=0, 0.3, 0.5 0.8 and 1 at 4.5 (7 ) K at Q=1.27 (2.64) Å$^{-1}$ (from ref. [82]). The open squares in (d-e) are from Ei=100 meV and open circles in (e) are from Ei=40 meV measured on MERLIN. The broad peak centred near 50 meV is the response across the hybridized bands, which is in agreement with the 55 meV charge gap observed through optical conductivity [67, 68].

Fig. 20 (color online) Colour coded inelastic neutron scattering intensity of $CeOs_2Al_{10}$ at 5 K and 38 K measured with incident energy of $E_i$=25 meV on the MERLIN spectrometer. The phonon contribution was subtracted using the data of non-magnetic reference compound $LaOs_2Al_{10}$.

Fig.21 (color online) The fit parameters, susceptibility, linewidth and peak position versus temperature obtained from fitting the magnetic scattering intensity of $CeOs_2Al_{10}$. The closed circles represent the fit using an inelastic peak and open circles represent the fit using a quasi-elastic peak. The green solid line in (a) shows the measured dc-susceptibility of the polycrystalline sample of $CeOs_2Al_{10}$ and in (b) the dotted and solid line represents the fits using exponential and $T^2$ behavior respectively (see text). In (c) the solid line represents the power law simulation $\Delta(T)=\Delta_0 (1-T/T_N)^\beta$ with $\Delta_0$ =11 meV, $\beta$=0.1 and $T_N$=28.5 K and the filled green diamonds are the optical gap ($\Delta_{op}$) from ref. [66].

**Refeneces**

## Table-I

The crystal structure type, magnetic properties, effective magnetic moment ($\mu_{eff}$), paramagnetic Curie-Weiss temperature ($\theta_P$) and ordering temperature ($T_C/T_N$) of the magnetically ordered ternary Ce intermetallic compounds. For comparison, we have also presented the magnetic ordering temperature of isostructural Gd compounds.

| Compound | Structure type | TMO | $T_{N,C}$ [K] | $\mu_{eff}$ [$\mu_B$] | $\theta_p$ [K] | TMO, $T_{C,N}$ of homolog Gd compound | References |
|---|---|---|---|---|---|---|---|
| CeRh$_3$B$_2$ | hexagonal | F | 120.8 | 3 | -373 | F, 90 | [71,22] |
| CeCuGe | hexagonal | F | 10 | 2.52 | -7.92 | AF, 14 | [150,151] |
| CePdSb | hexagonal | F | 17 | 2.6 | 10 | F, 17 | [152] |
| CePtSb | hexagonal | F | 4.5 | 2.62 | -34.4 | - | [153] |
| CePdAs | hexagonal | F | 4 | 2.6 | -0.73 | - | [154][+] |
| CeCoGa | monoclinic | AF | 4.3 | 1.8 | -80.5 | - | [155] |
| CeRuSn | monoclinic | AF | 2.8 | 2.1 | -1.9 | - | [156] |
| CePdIn | hexagonal | AF | 1.8 | 2.56 | -15 | - | [157] |
| CePdSn | orthorhombic | AF | 7 | 2.50 | -40 | - | [158] |
| CePtSn | orthorhombic | AF | 7 | 2.7 | -40 | - | [159] |
| CeNiC$_2$ | orthorhombic | AF | 20,10,2.2 | 2.47 | -18.3 | AF, 20.0 | [160] |
| CeRuSi$_2$ | monoclinic | F | 11.7 | 1.7 | -40 | - | [161] |
| CeNiGe$_2$ | orthorhombic | AF | 3.2,3.9 | 2.5 | -20.8 | AF, 24.5 | [162] |
| CePdSb$_3$ | orthorhombic | AF | 3.1 | 2.54 | - | - | [163] |
| CeCoGe$_3$ | tetragonal | AF | 21.0 | 2.43 | -30.4 | | [179] |
| CeRu$_2$Ge$_2$ | tetragonal | F | 8 | 2.83 | 40.3 | AF, 32, 29 | [164] |
| CeIr$_2$B$_2$ | orthorhombic | F | 5.1 | 2.5 | -13.4 | - | [165] |
| CeRh$_2$Si$_2$ | tetragonal | AF | 36 | 2.9 | 72 | AF, 106 | [166] |
| CeRu$_2$Al$_{10}$ | orthorhombic | AF | 27 | 3.03 | -44 | AF, 16 | [59] |
| CeOs$_2$Al$_{10}$ | orthorhombic | AF | 29 | 2.7 | -30 | AF, 18 | [61] |
| Ce$_5$Rh$_4$Sn$_{10}$ | tetragonal | AF | 4.4 | 2.34 | -8.04 | - | [167] |
| Ce$_5$Ir$_4$Sn$_{10}$ | tetragonal | AF | 4.2 | 2.09 | -11.88 | - | [167] |
| CeRh$_2$Sn$_4$ | orthorhombic | AF | 3.2 | 2.47 | -22 | - | [168] |
| Ce$_3$Rh$_4$Sn$_{13}$ | cubic | AF | 1.2,2 | 2.53 | -29 | - | [169][+] |
| Ce$_3$Pt$_4$In$_{13}$ | cubic | AF | 0.95 | 2.64 | -36 | - | [170] |
| Ce$_2$NiSi$_3$ | hexagonal | AF | 3.2 | 2.56 | -10 | - | [171] |
| Ce$_4$Ni$_6$Al$_{23}$ | monoclinic | AF,F | 3, 6 | 2.60 | -225 | - | [172] |
| CeRh$_2$Si | orthorhombic | AF | 1.65 | 2.45 | -65 | - | [173] |
| CePt$_3$Si | tetragonal | AF | 2.2 | 2.54 | -46 | AF, 15.1 | [174] |
| CeOs$_4$Sb$_{12}$ | cubic | AF | 1.6 | 2.41 | -28 | - | [175] |
| Ce$_2$RuZn$_4$ | tetragonal | AF | 2 | 2.57 | -2.6 | - | [176] |
| Ce$_2$Ga$_{12}$Pt | tetragonal | AF | 7.3, 5.5 | 2.46 | -20 | - | [177] |
| CeScSi | Tetragonal | AF | 26.0 | 2.59 | 15.7 | | [178] |
| CeScGe | tetragonal | AF | 46.0 | | | | [178] |

*TMO: type of magnetic ordering; AF: antiferromagnetic; F: ferromagnetic

+ Single crystal data

Table II. Lattice parameters, unit cell volume and selected interatomic distances for $CeT_2Al_{10}$ (T=Fe, Ru and Os) at 300K.

| Compounds | a (Å) | b (Å) | c (Å) | V (Å$^3$) | Ce-Ce (Å) | Ce-T (Å) | Ce-Al(Å) (1,2,3) |
|---|---|---|---|---|---|---|---|
| $CeFe_2Al_{10}$ | 9.0159 | 10.2419 | 9.0882 | 839.204 | 5.2032 | 3.4467 | 3.1726 3.1545 3,2155 |
| $CeRu_2Al_{10}$ | 9.1320 | 10.2871 | 9.1933 | 863.635 | 5.2604 | 3.4828 | 3.1971 3.1947 3.2486 |
| $CeOs_2Al_{10}$ | 9.1386 | 10.2662 | 9.1852 | 861.744 | 5.2500 | 3.4711 | 3.1469 3.1867 3.2324 |

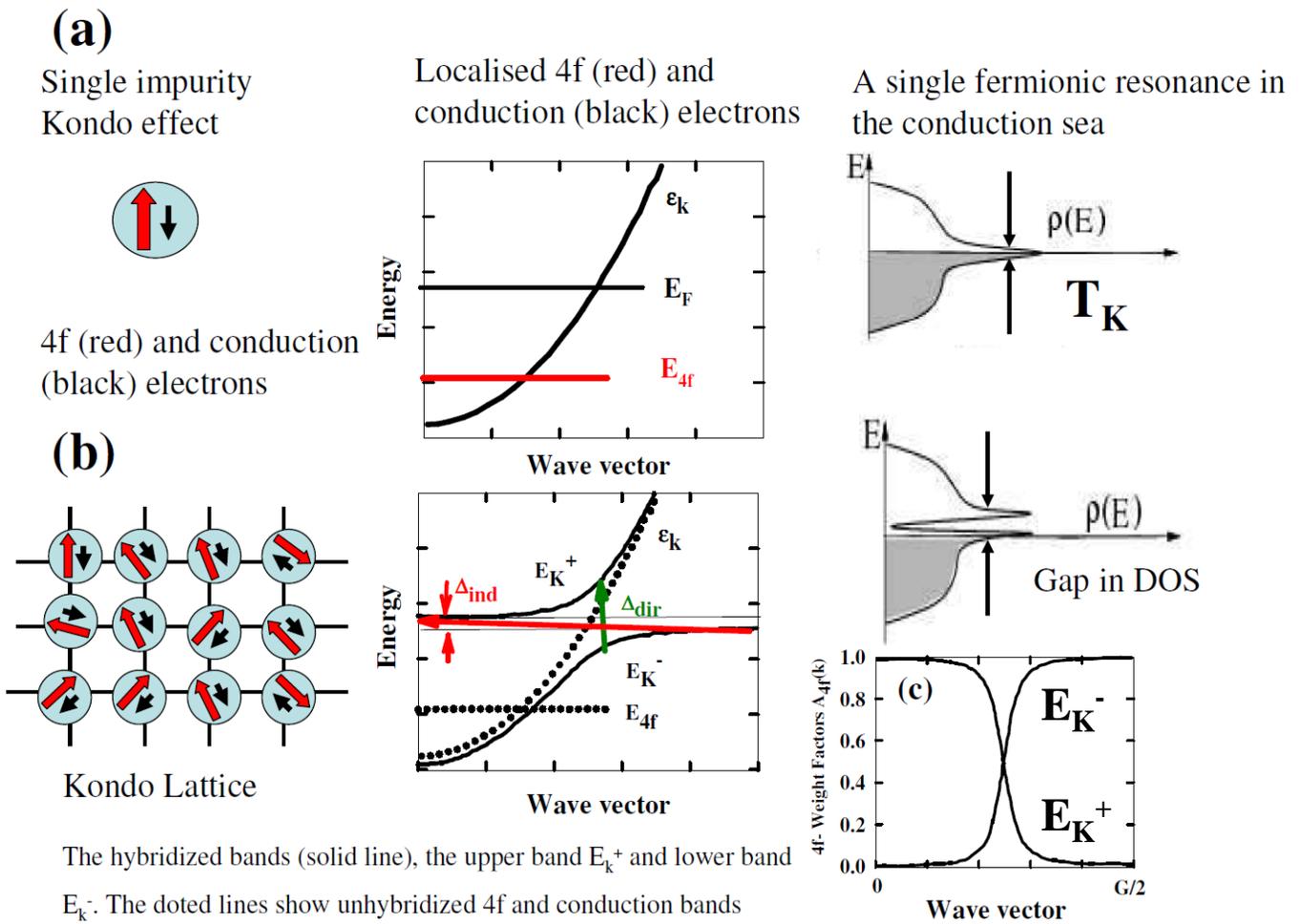

The hybridized bands (solid line), the upper band $E_k^+$ and lower band $E_k^-$. The doted lines show unhybridized 4f and conduction bands

Fig.1 Adroja et al

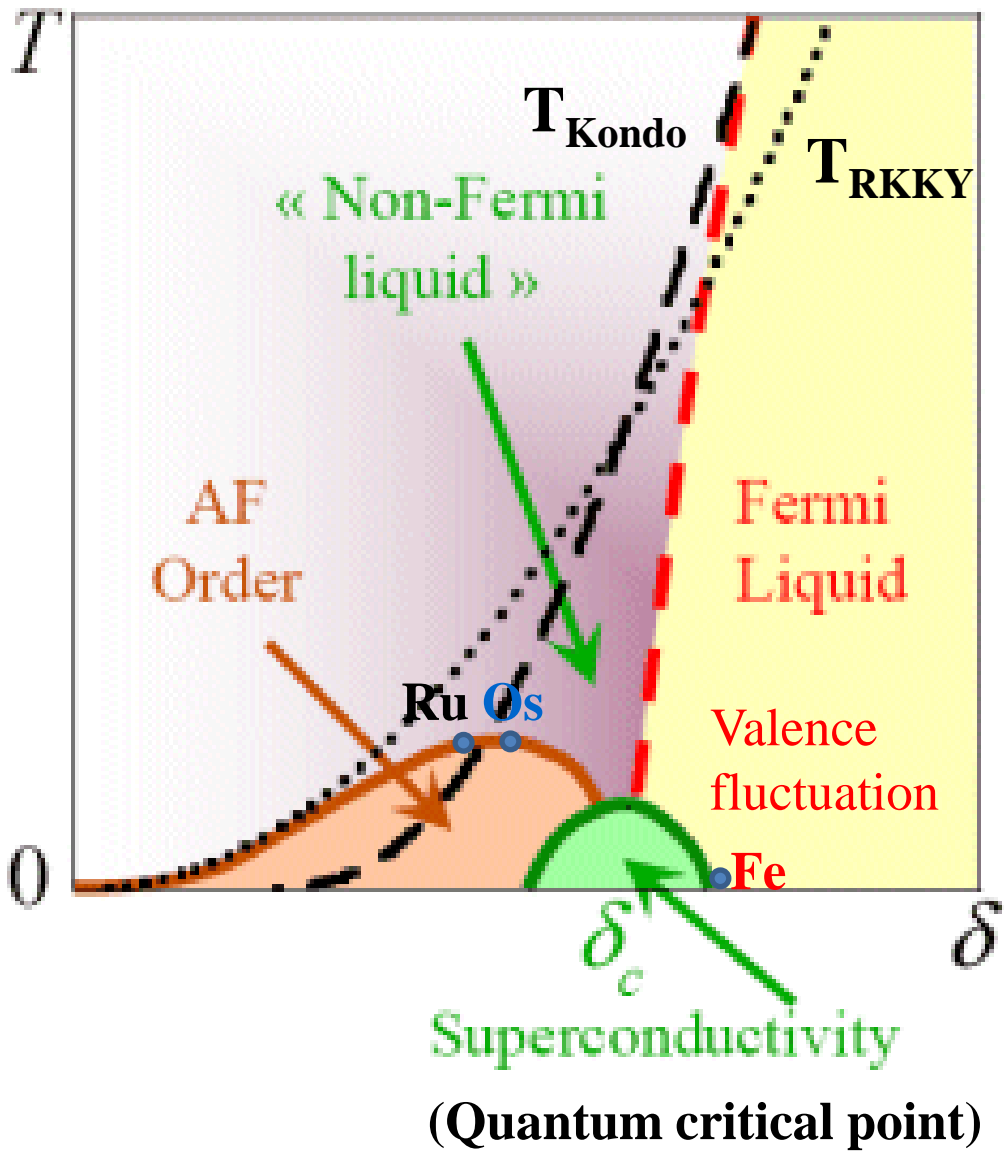

Fig.2 Adroja et al

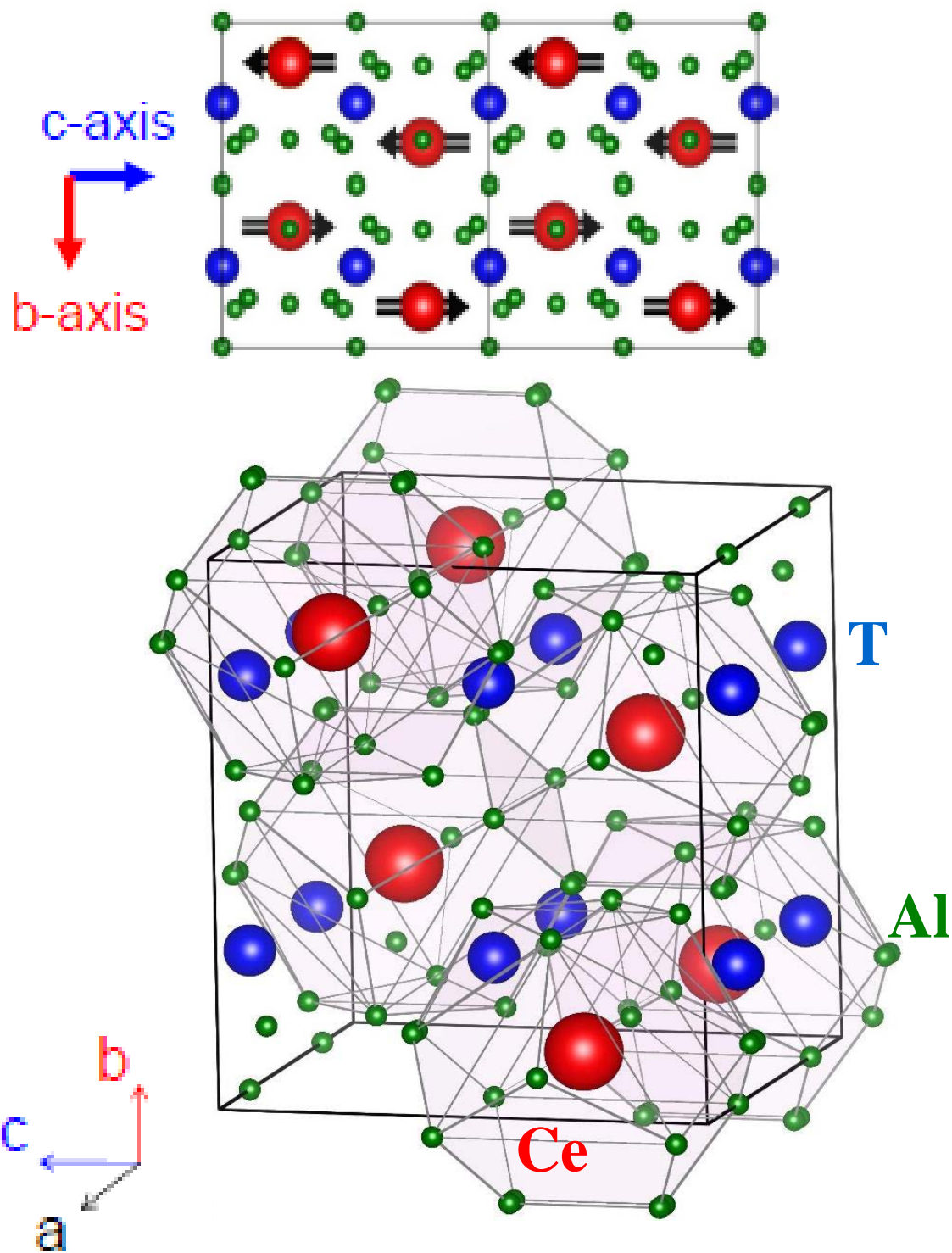

Fig.3 Adroja et al

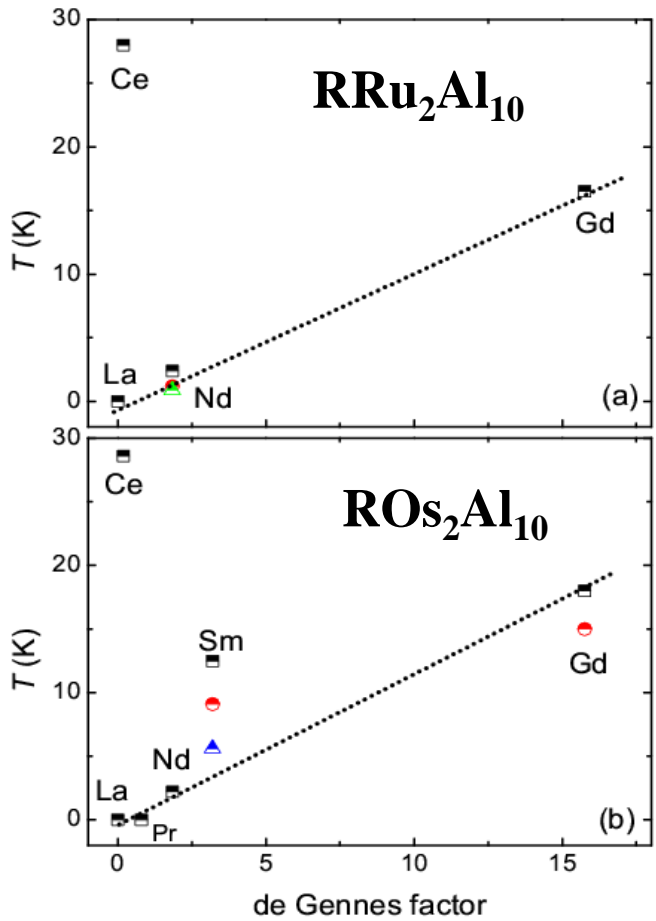

Fig.4 Adroja et al

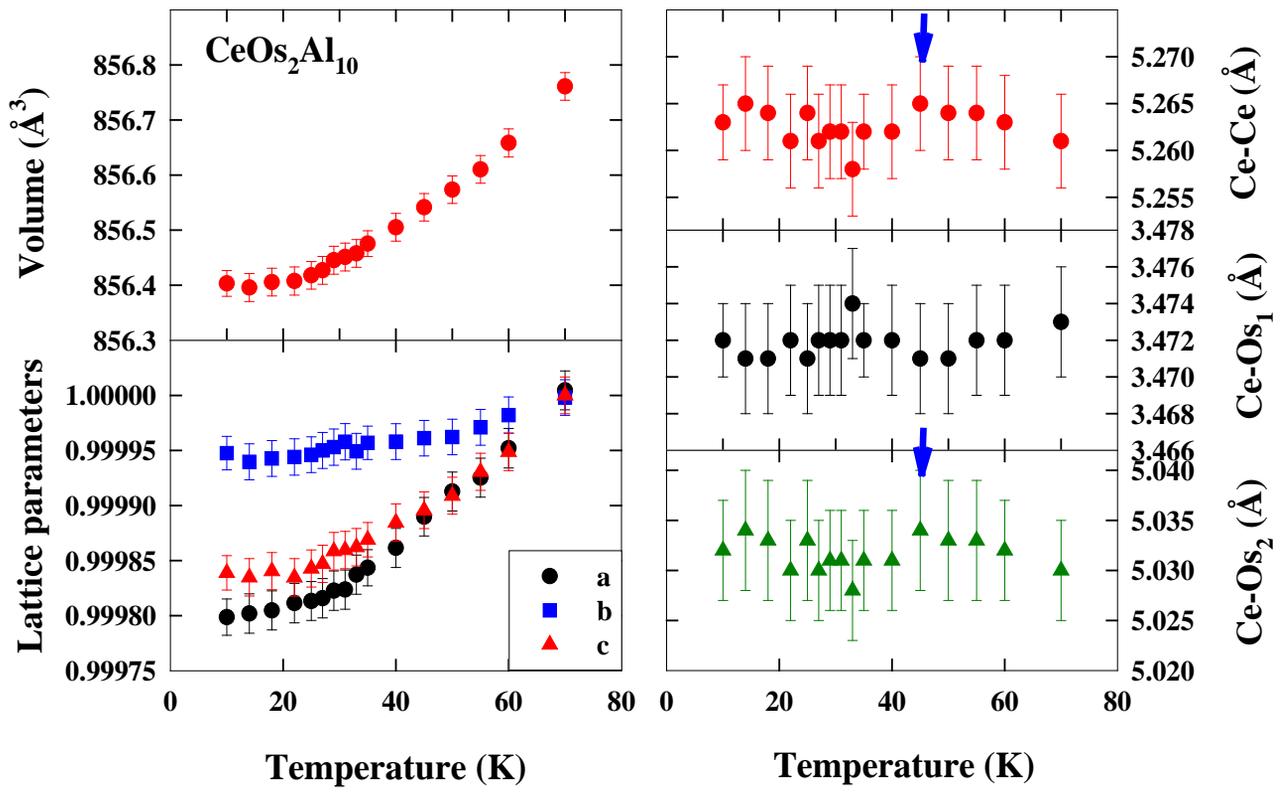

Fig. 5 Adroja et al

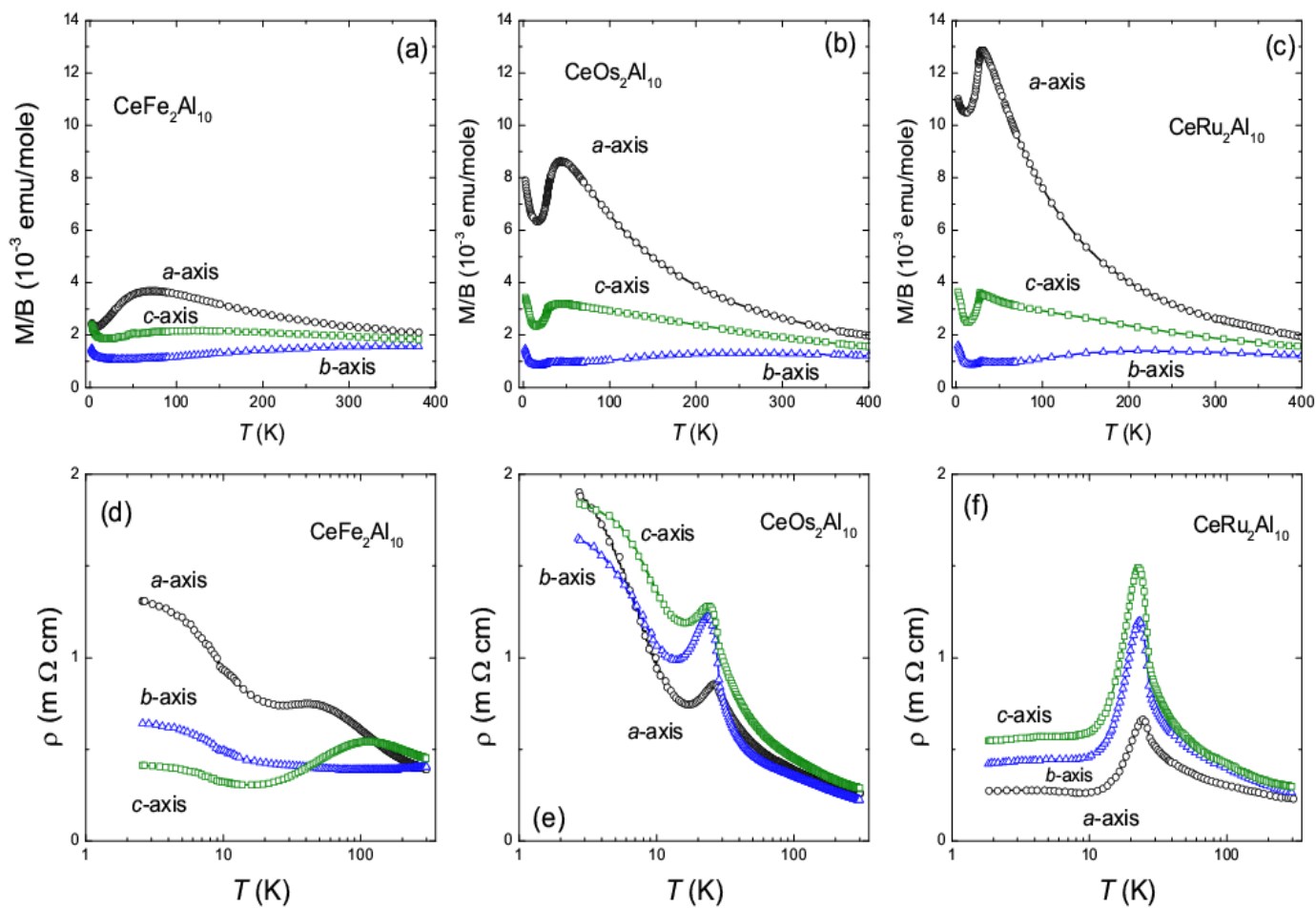

Fig. 6 Adroja et al

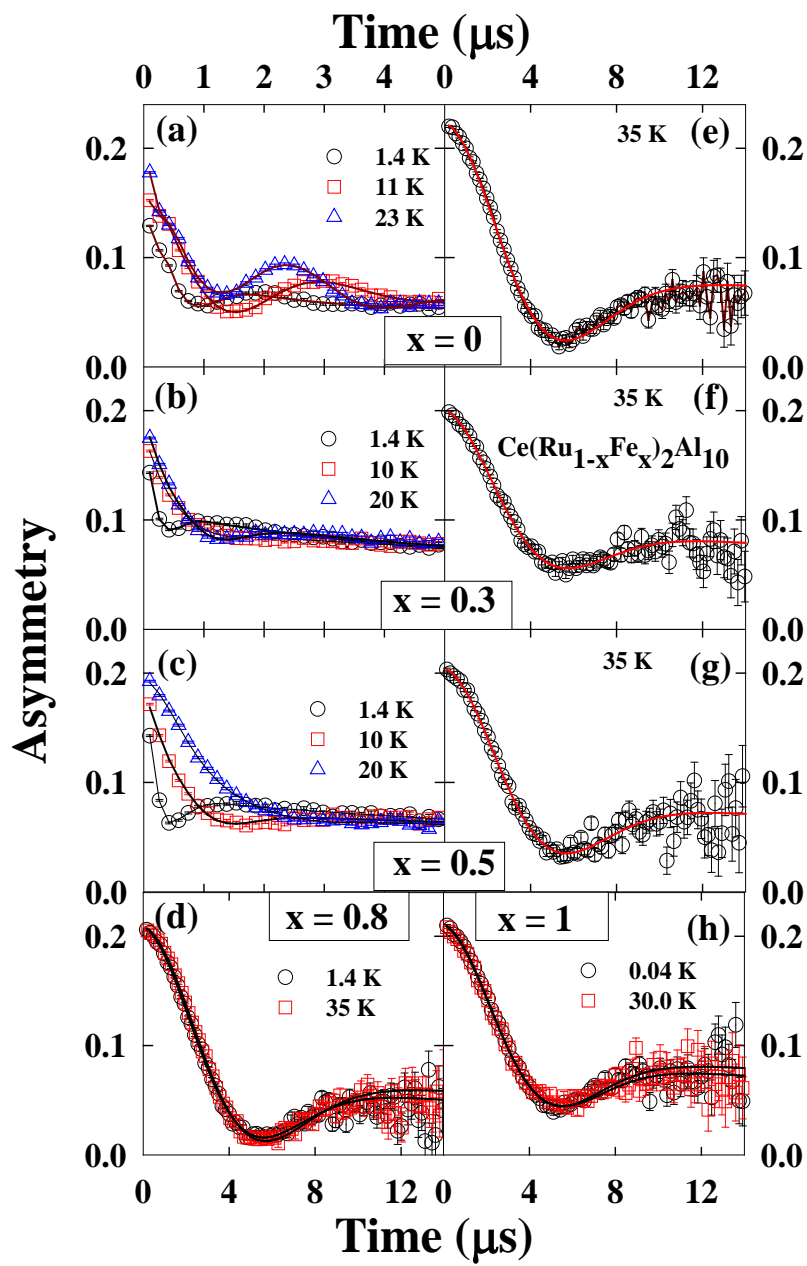

Fig. 7 Adroja et al

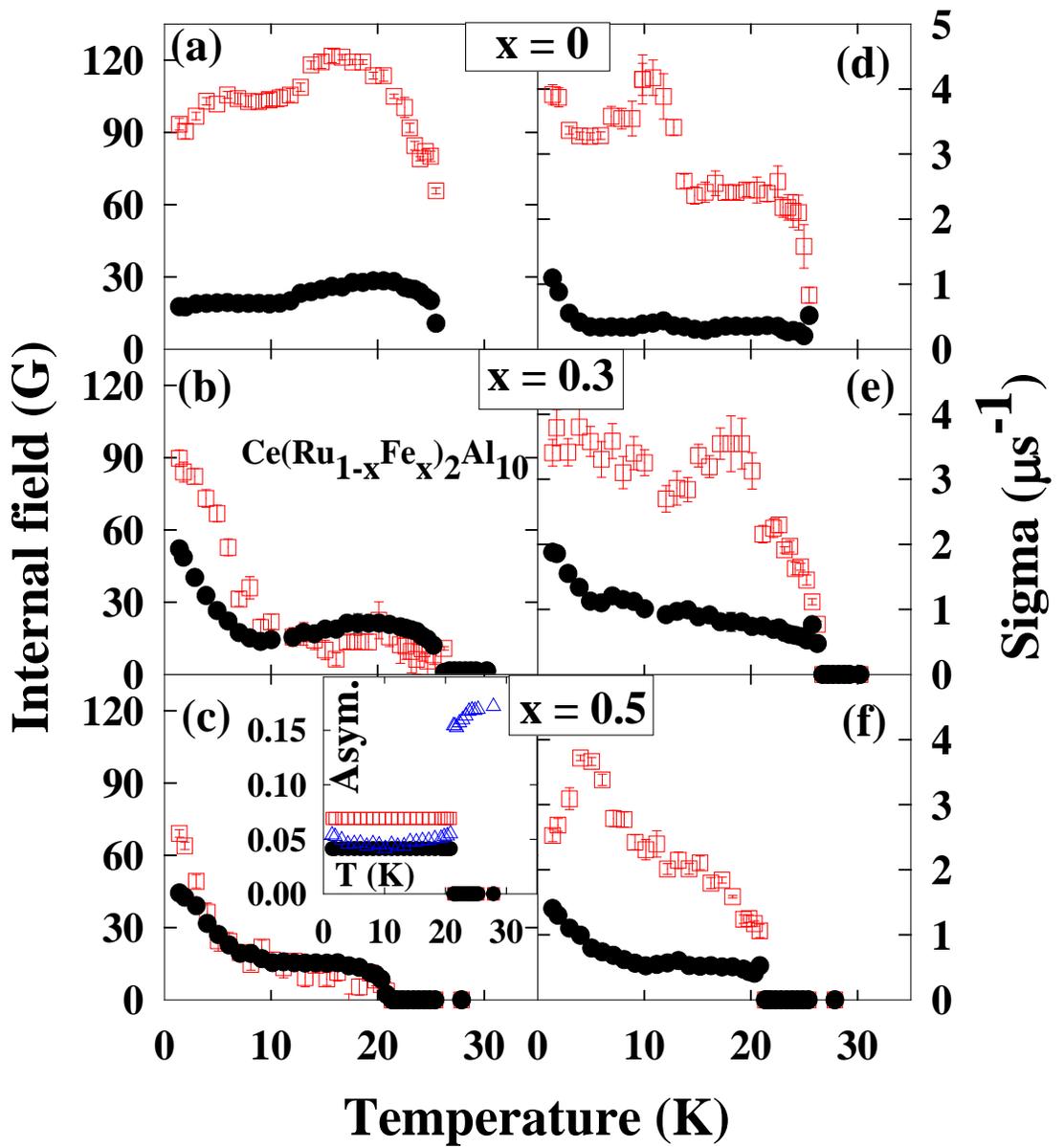

Fig.8 Adroja et al

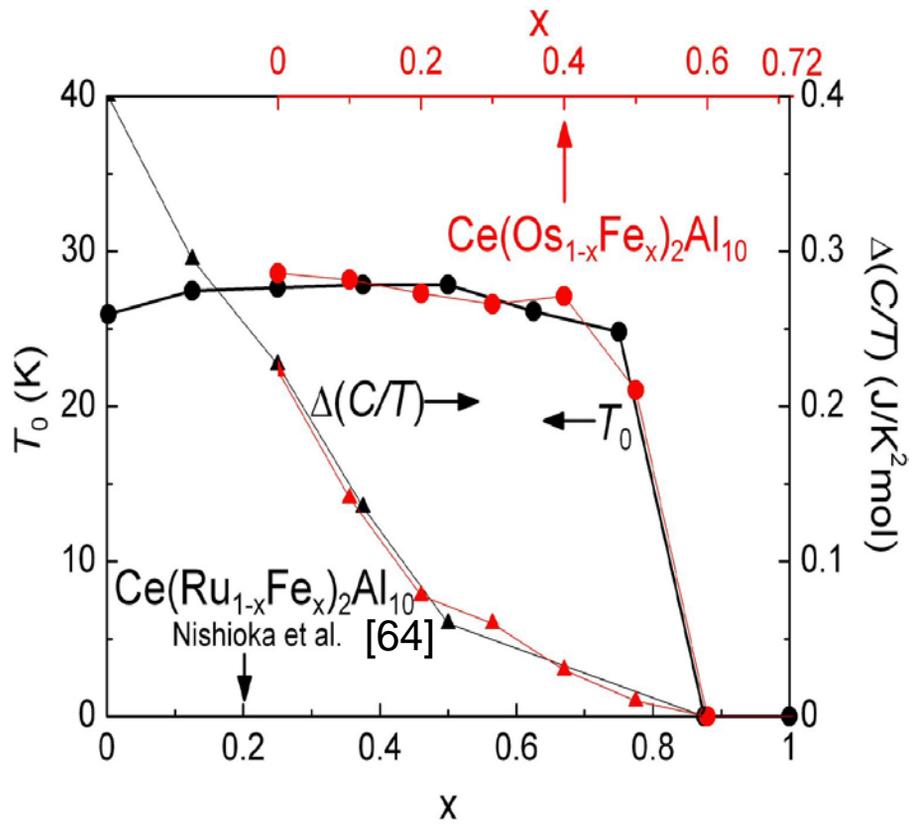

Fig. 9 Adroja et al

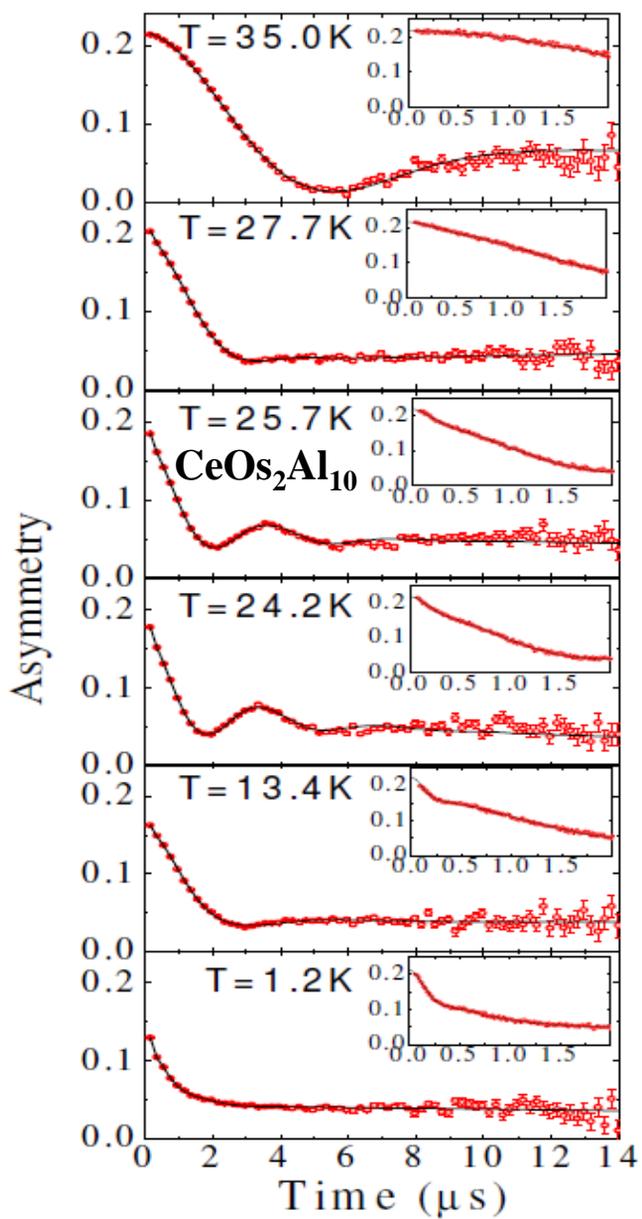

Fig. 10 Adroja et al

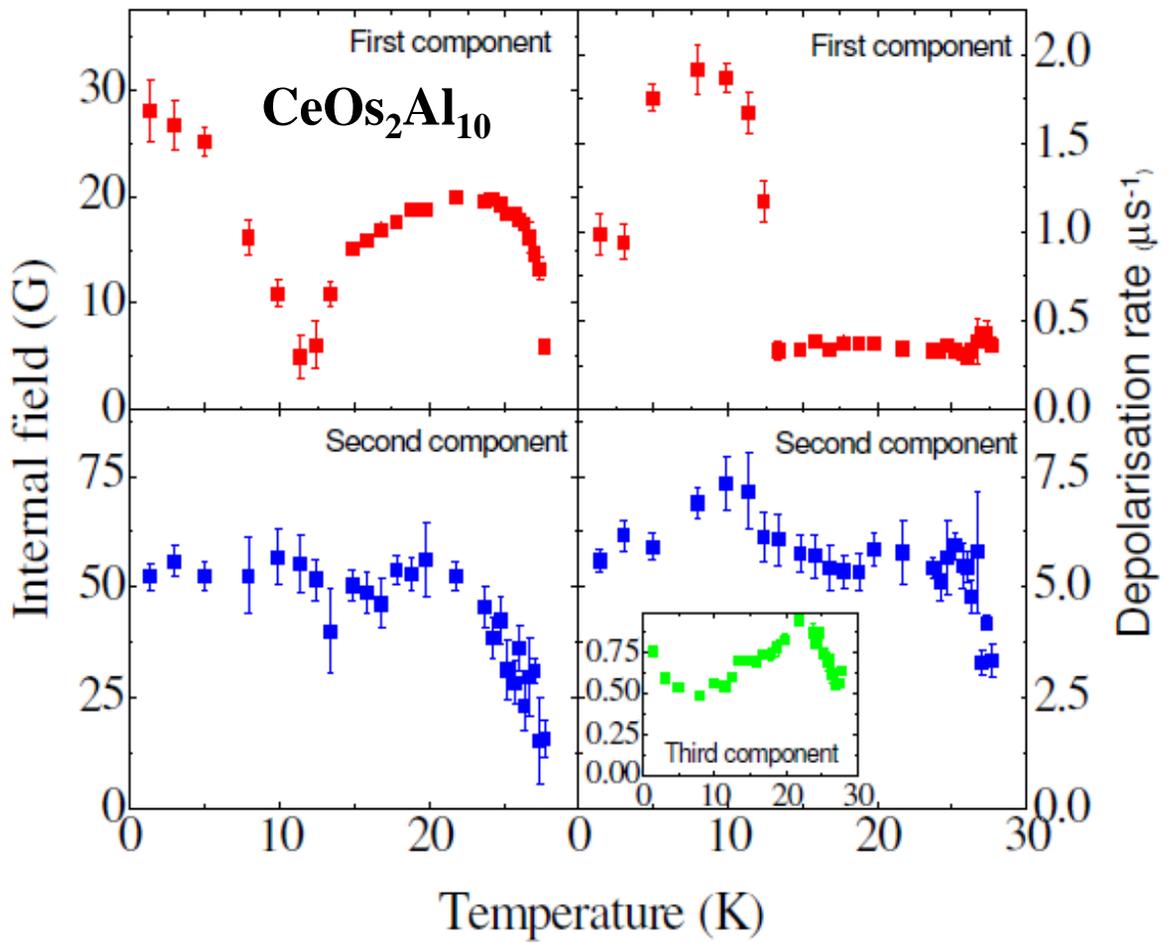

Fig. 11 Adroja et al

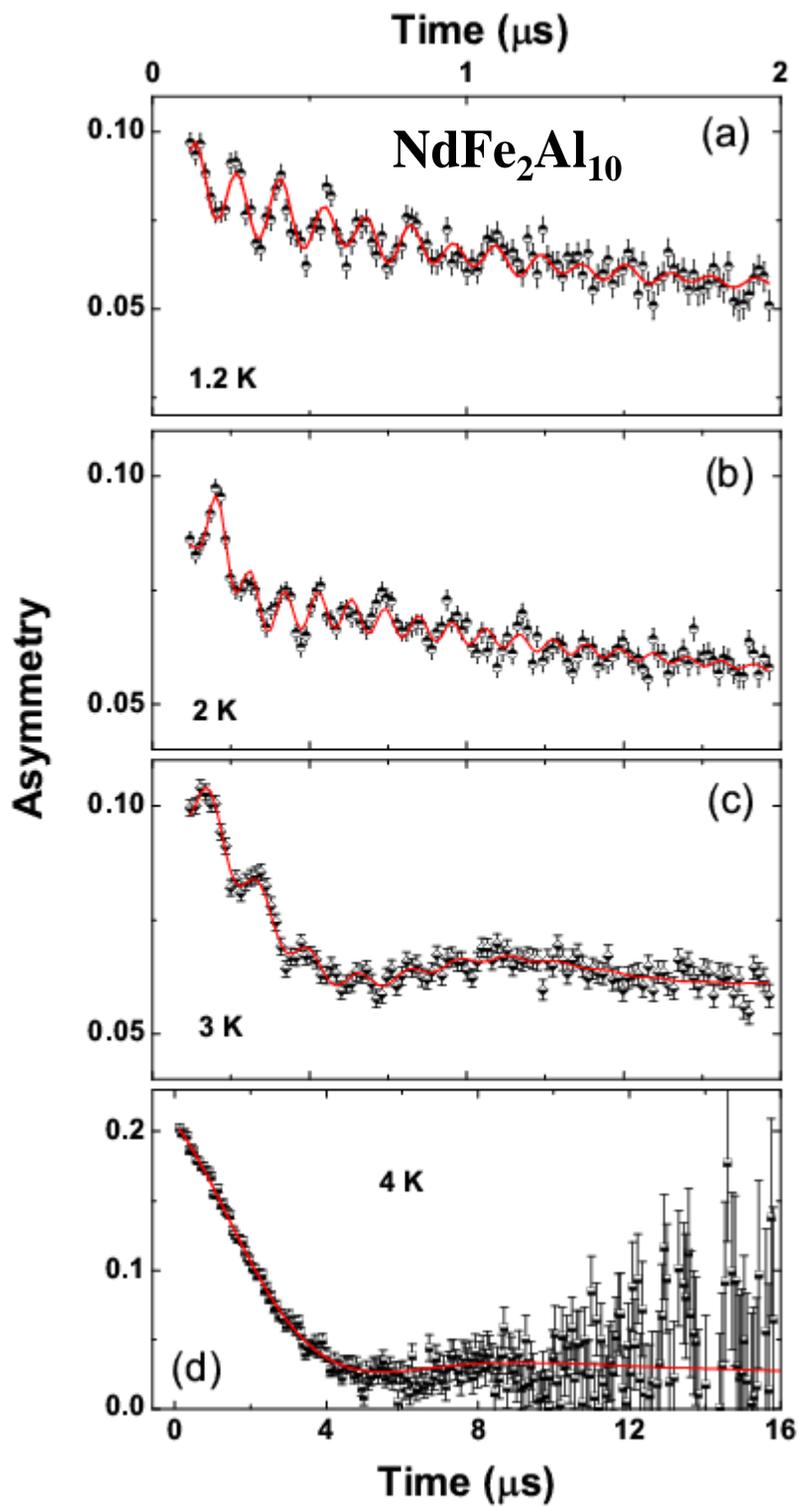

Fig. 12 Adroja et al

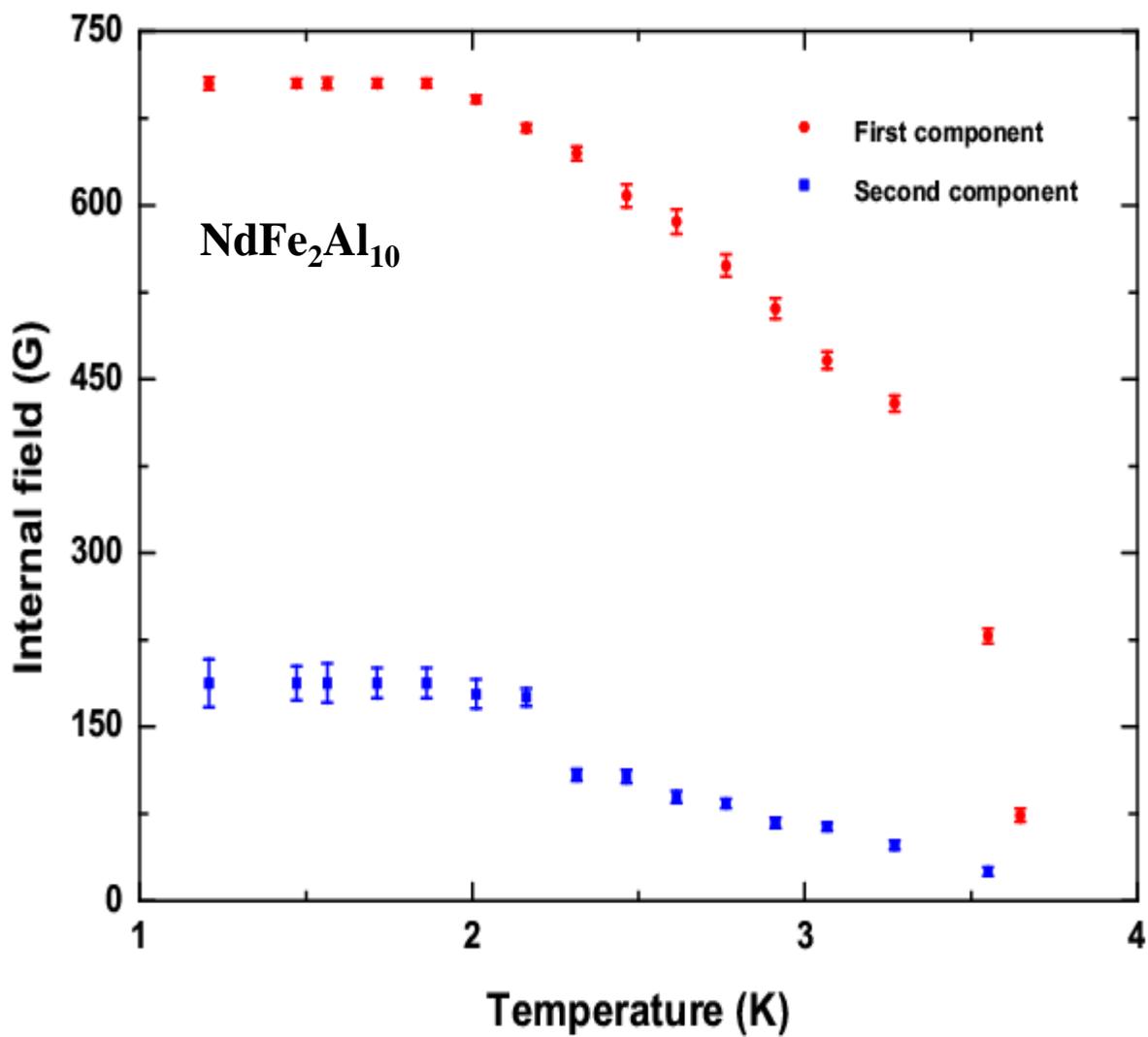

Fig. 13 Adroja et al

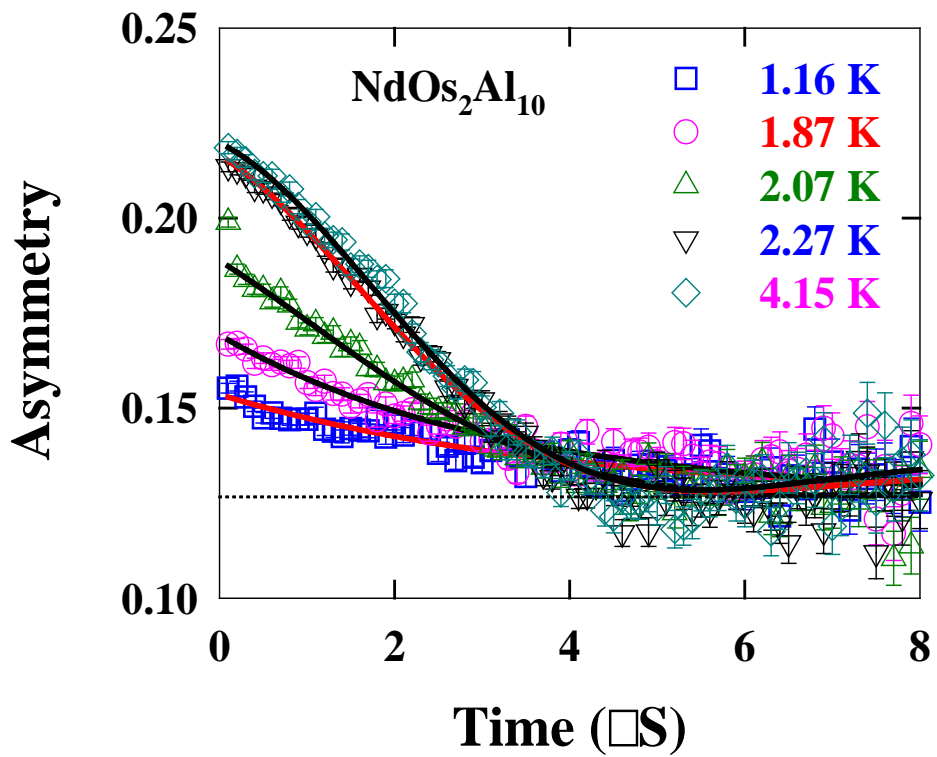

Fig. 14 Adroja et al

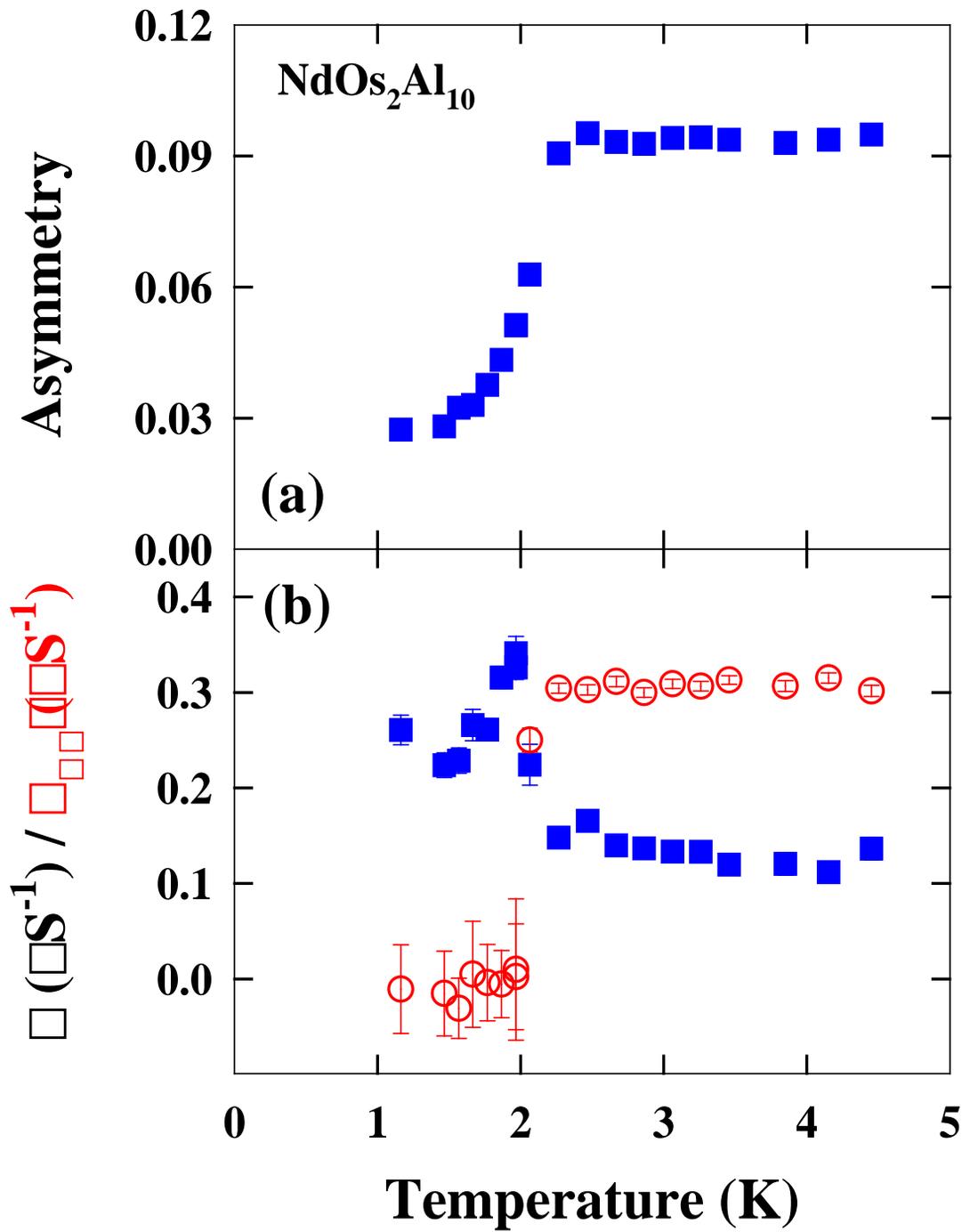

Fig.15 Adroja et al

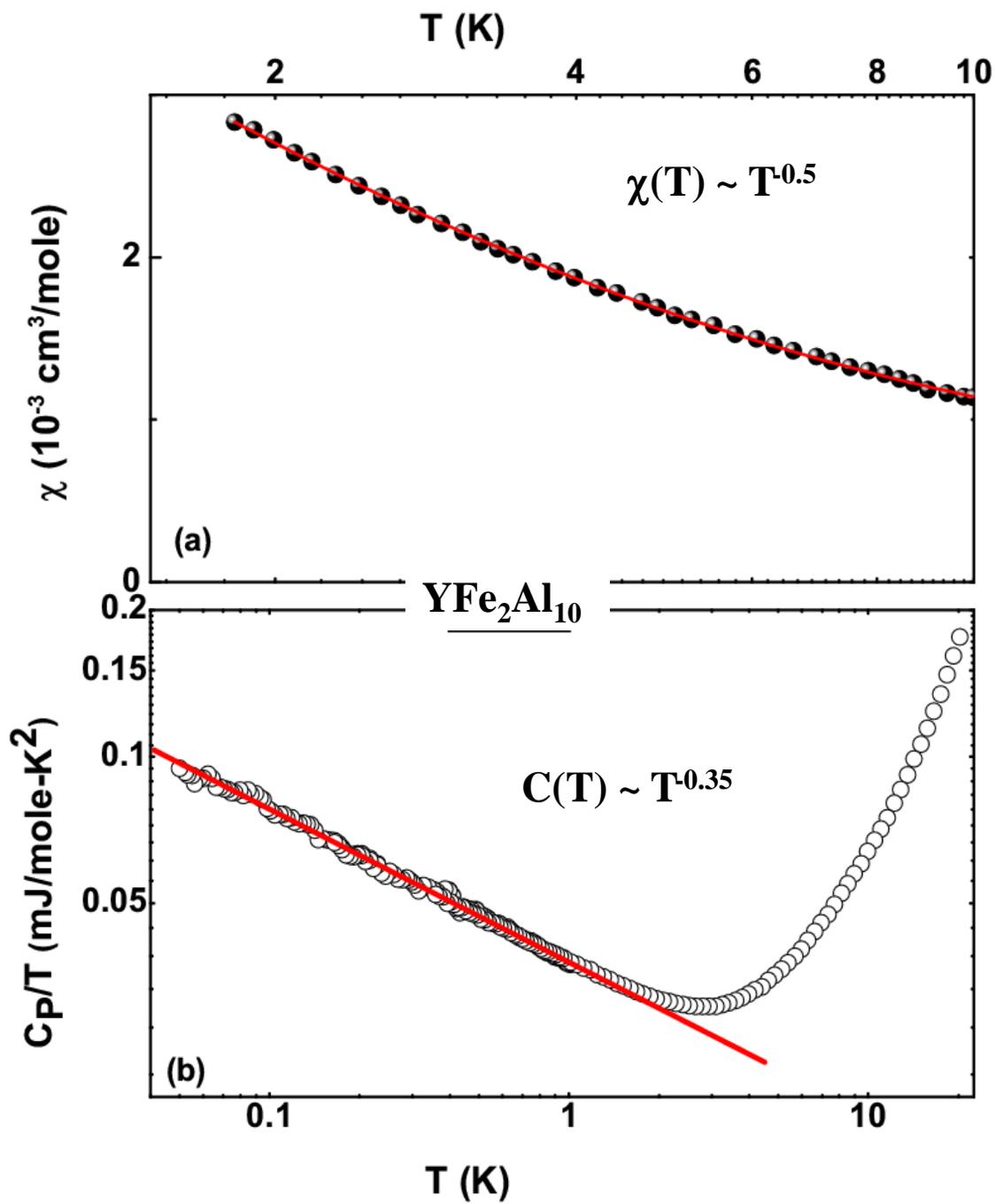

Fig. 16 Adroja et al

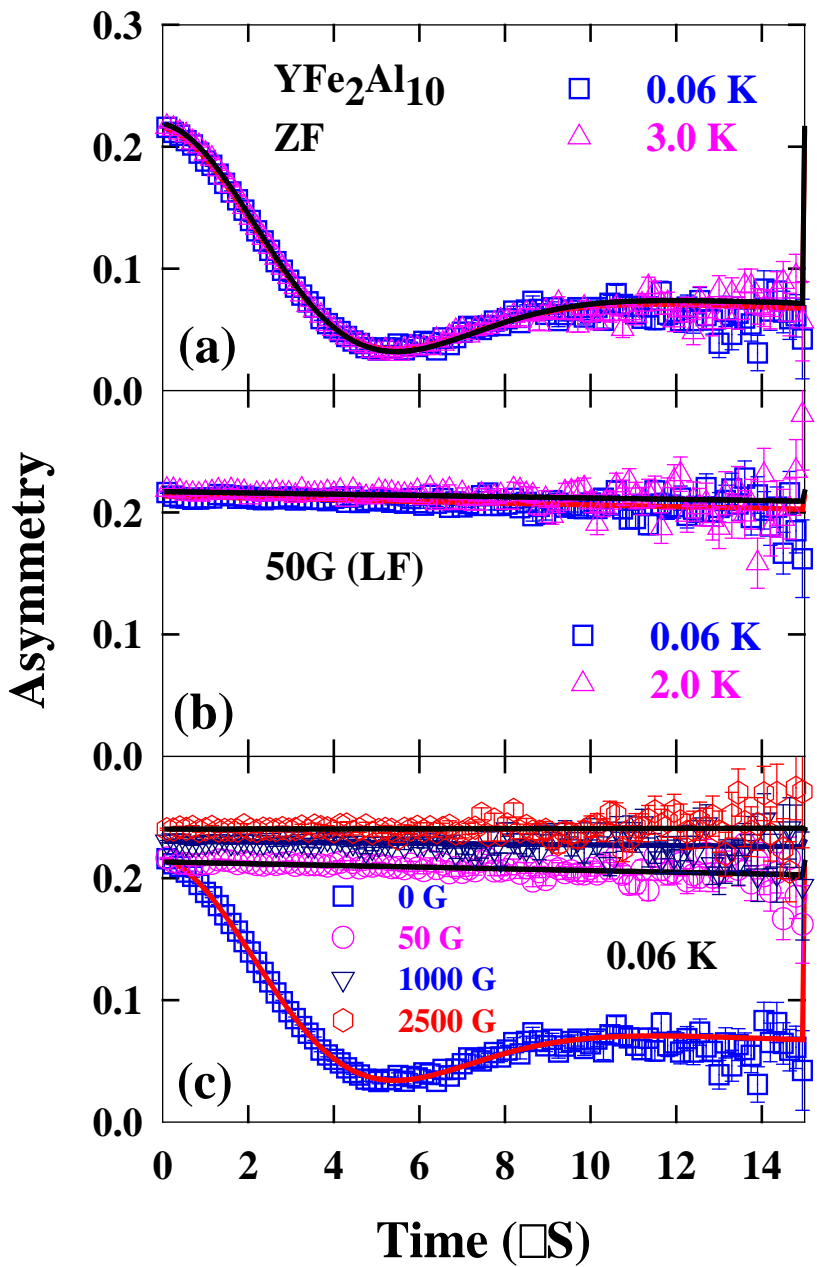

Fig. 17 Adroja et al

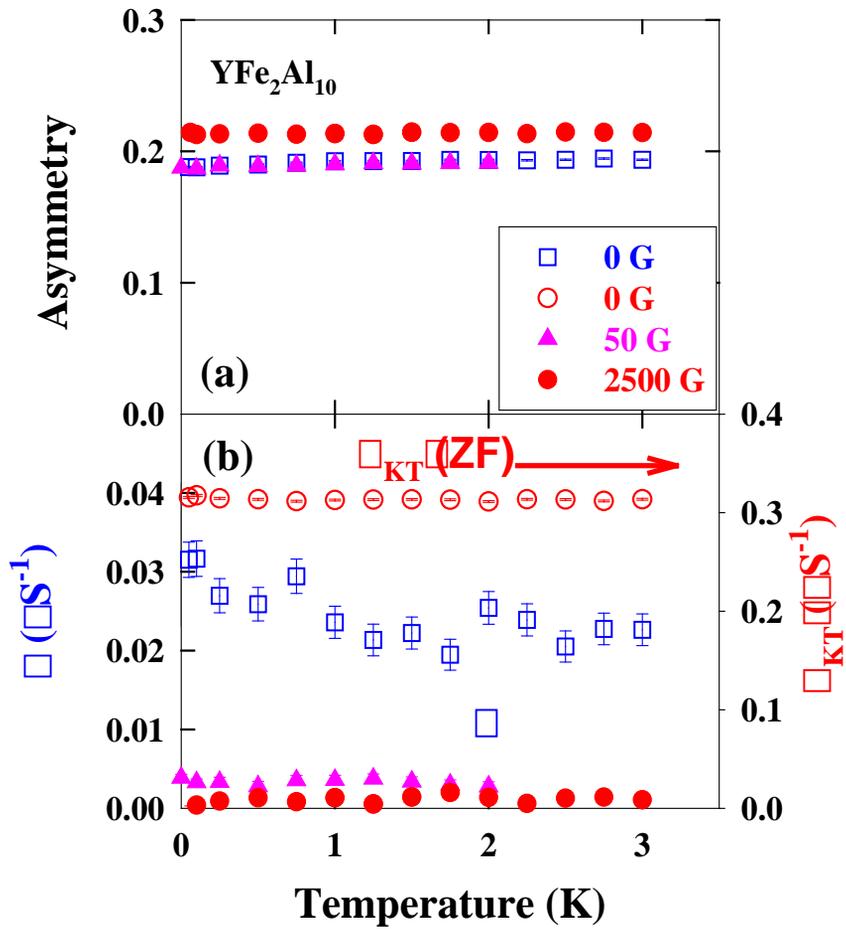

Fig. 18 Adroja et al

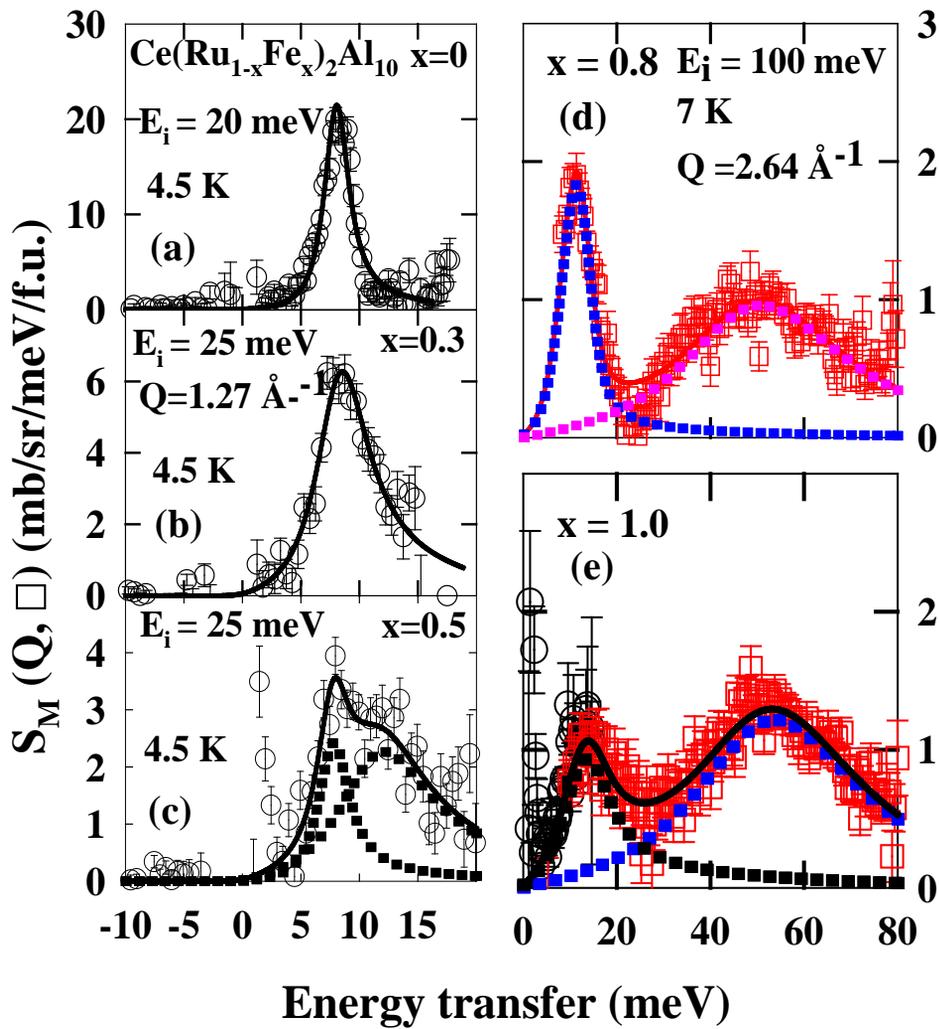

Fig. 19 Adroja et al

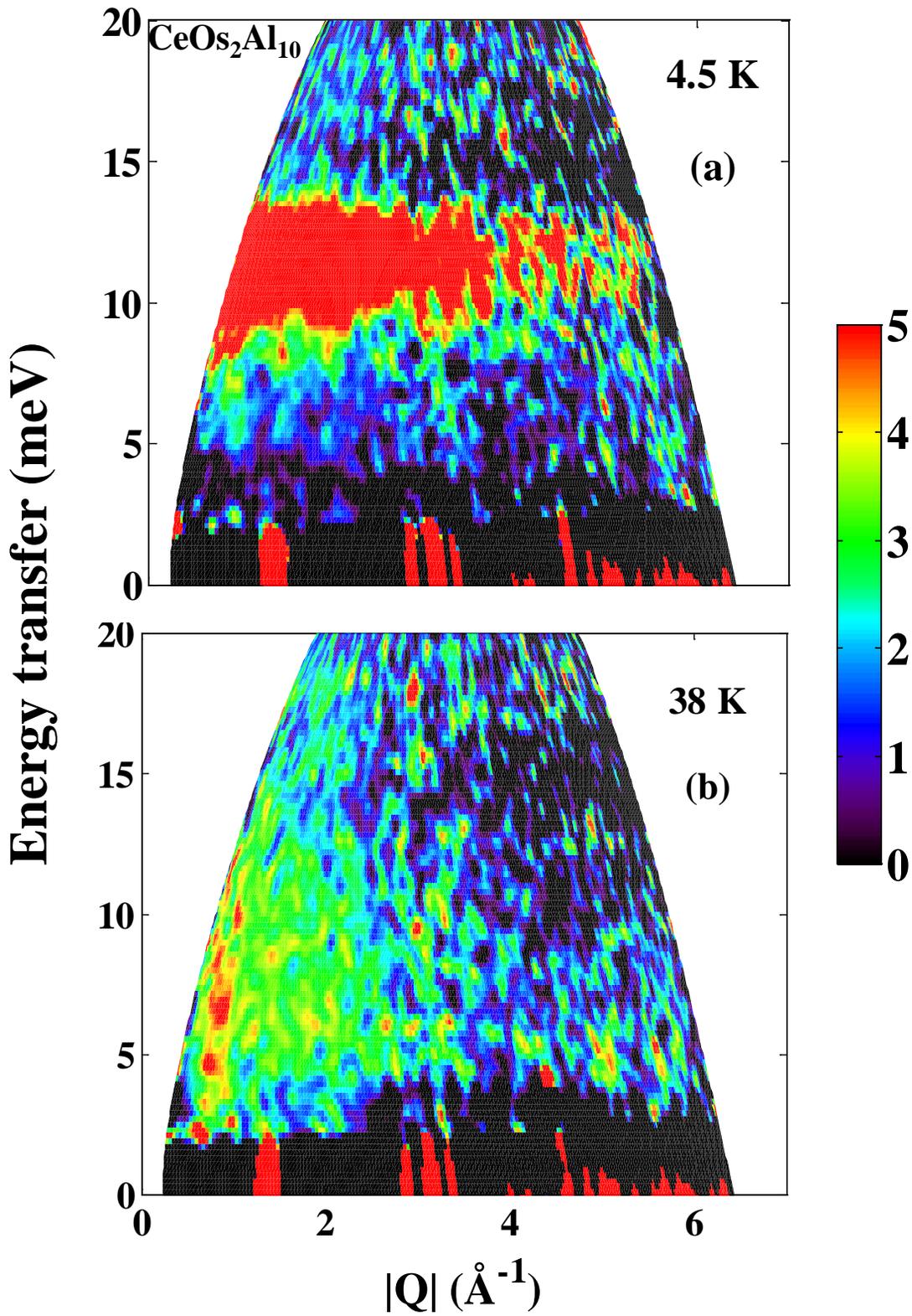

Fig. 20 Adroja et al

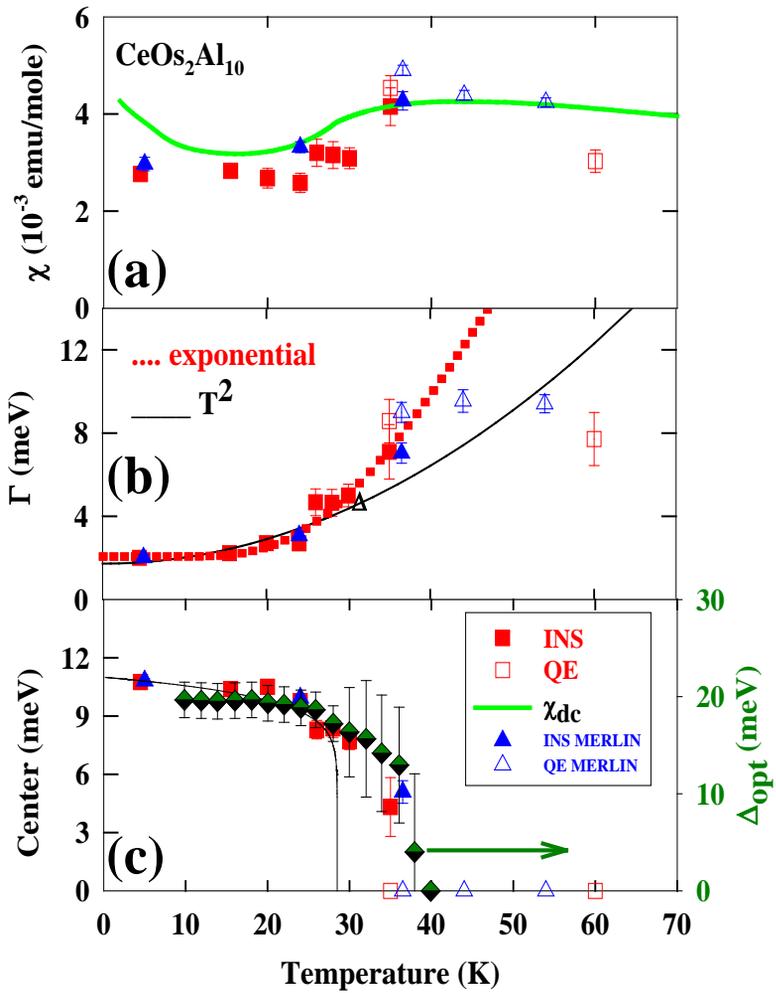

Fig. 21 Adroja et al